\documentclass[10pt,twocolumn,letterpaper]{article}

\usepackage{titling}
\usepackage{iccv}
\usepackage{times}
\usepackage{epsfig}
\usepackage{graphicx}
\usepackage{amsmath}
\usepackage{amssymb}
\usepackage{comment}
\usepackage{color}
\usepackage{threeparttable}
\usepackage{multirow}
\usepackage{multicol}
\usepackage{setspace}
\usepackage{adjustbox}
\usepackage{bm}
\usepackage{subfig}
\usepackage{mathrsfs}
\usepackage{cuted}
\usepackage{capt-of}
\usepackage{tabu}
\usepackage{tabularx}
\usepackage{xcolor}

\definecolor{dodgerblue}{rgb}{0.12, 0.56, 1.0}
\definecolor{limegreen}{rgb}{0.2, 0.8, 0.2}
\definecolor{peru}{rgb}{0.8,0.52,0.25}
\newcolumntype{C}{>{\centering\arraybackslash}X}

\makeatletter
\newcommand{\printfnsymbol}[1]{%
	\textsuperscript{\@fnsymbol{#1}}%
}
\makeatother

\usepackage[pagebackref=true,breaklinks=true,letterpaper=true,colorlinks,bookmarks=false]{hyperref}

\iccvfinalcopy 

\def\iccvPaperID{3013} 

\ificcvfinal\fi

\begin{document}

\title{Best-Buddy GANs for Highly Detailed Image Super-Resolution}


\author{
	Wenbo Li\textsuperscript{1}\thanks{Equal contribution} \quad Kun Zhou\textsuperscript{2}\printfnsymbol{1} \quad Lu Qi\textsuperscript{1} \quad Liying Lu\textsuperscript{1} \quad Nianjuan Jiang\textsuperscript{2} \quad Jiangbo Lu\textsuperscript{2}\thanks{Corresponding author. Email: jiangbo@smartmore.com} \quad Jiaya Jia\textsuperscript{1,2} \\
	$^{1}$The Chinese University of Hong Kong \quad ${^2}$Smartmore Corporation\\
	{\tt\small\{wenboli,luqi,lylu\}@cse.cuhk.edu.hk} \\
	{\tt\small\{kun.zhou,nianjuan.jiang,jiangbo\}@smartmore.com} \\
}

\maketitle
\ificcvfinal\thispagestyle{empty}\fi

\begin{strip}\centering
	\includegraphics[width=1.0\linewidth]{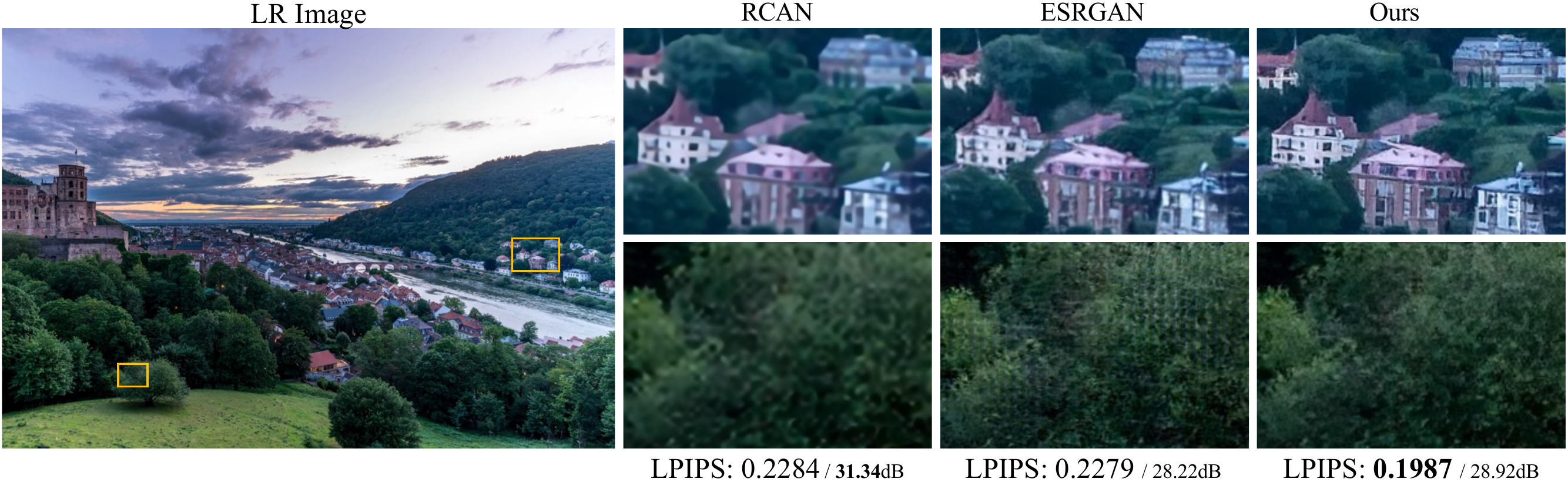}
	\captionof{figure}{Visual comparison of our best-buddy GANs, PSNR-oriented RCAN~\cite{zhang2018image} and perception-oriented ESRGAN~\cite{wang2018esrgan}. The numbers are LPIPS($\downarrow$)/PSNR($\uparrow$). Richer details are generated in our result with much fewer artifacts.\label{fig:teasing}} 
\end{strip}

\begin{abstract}
	We consider the single image super-resolution (SISR) problem, where a high-resolution (HR) image is generated based on a low-resolution (LR) input. Recently, generative adversarial networks (GANs) become popular to hallucinate details. Most methods along this line rely on a predefined single-LR-single-HR mapping, which is not flexible enough for the SISR task. Also, GAN-generated fake details may often undermine the realism of the whole image. We address these issues by proposing best-buddy GANs (Beby-GAN) for rich-detail SISR. Relaxing the immutable one-to-one constraint, we allow the estimated patches to dynamically seek the best supervision during training, which is beneficial to producing more reasonable details. Besides, we propose a region-aware adversarial learning strategy that directs our model to focus on generating details for textured areas adaptively. Extensive experiments justify the effectiveness of our method. An ultra-high-resolution 4K dataset is also constructed to facilitate future super-resolution research.\footnote[1]{Please find a higher quality version of this paper at \href{https://github.com/Jia-Research-Lab/Simple-SR}{https://github.com/Jia-Research-Lab/Simple-SR}.}
\end{abstract}

\section{Introduction}

The increasing demand for high-quality displays has promoted the rapid development of single image super-resolution (SISR). SISR has been successfully applied to a wide range of tasks, such as medical diagnostic imaging, security imaging and satellite imaging.

A great number of methods were proposed based on insightful image priors and optimization techniques, such as self-similarity~\cite{protter2008generalizing,glasner2009super,yang2010exploiting,huang2015single} and sparsity~\cite{martin2001database,yang2008image,yang2010image,zeyde2010single,timofte2013anchored,romano2014single,peleg2014statistical}. In recent years, deep-learning-based methods~\cite{dong2014learning,kim2016accurate,kim2016deeply,shi2016real,lai2017deep,tai2017image,tai2017memnet,lim2017enhanced,haris2018deep,zhang2018residual,zhang2018image} further advanced SISR. Most of them rely on an immutable one-to-one supervision to pursue high PSNR but possibly generate blurry results. For example, the solution of commonly adopted one-to-one MSE/MAE metric approximates mean or median of data~\cite{sonderby2016amortised}. As shown in Figure~\ref{fig:teasing}, the HR estimation of RCAN~\cite{zhang2018image} achieves the highest PSNR, yet lacking high-frequency texture.

 To enhance the perceptual quality of recovered images, several methods~\cite{johnson2016perceptual,ledig2017photo,mechrez2018maintaining,wang2018esrgan,zhang2019ranksrgan,soh2019natural} use adversarial learning and perceptual loss~\cite{johnson2016perceptual}. It is noted that the issue of excessive smoothing caused by the one-to-one MSE/MAE loss is still not optimally addressed. Besides, the training of generative adversarial networks (GANs)~\cite{goodfellow2014generative} could be unstable and result in unpleasant visual artifacts (see the recovered tree of ESRGAN~\cite{wang2018esrgan} in Figure~\ref{fig:teasing}). Thus in this paper, we aim to address these issues from two aspects.

\begin{figure}[t]
	\begin{center}
		\includegraphics[width=0.95\linewidth]{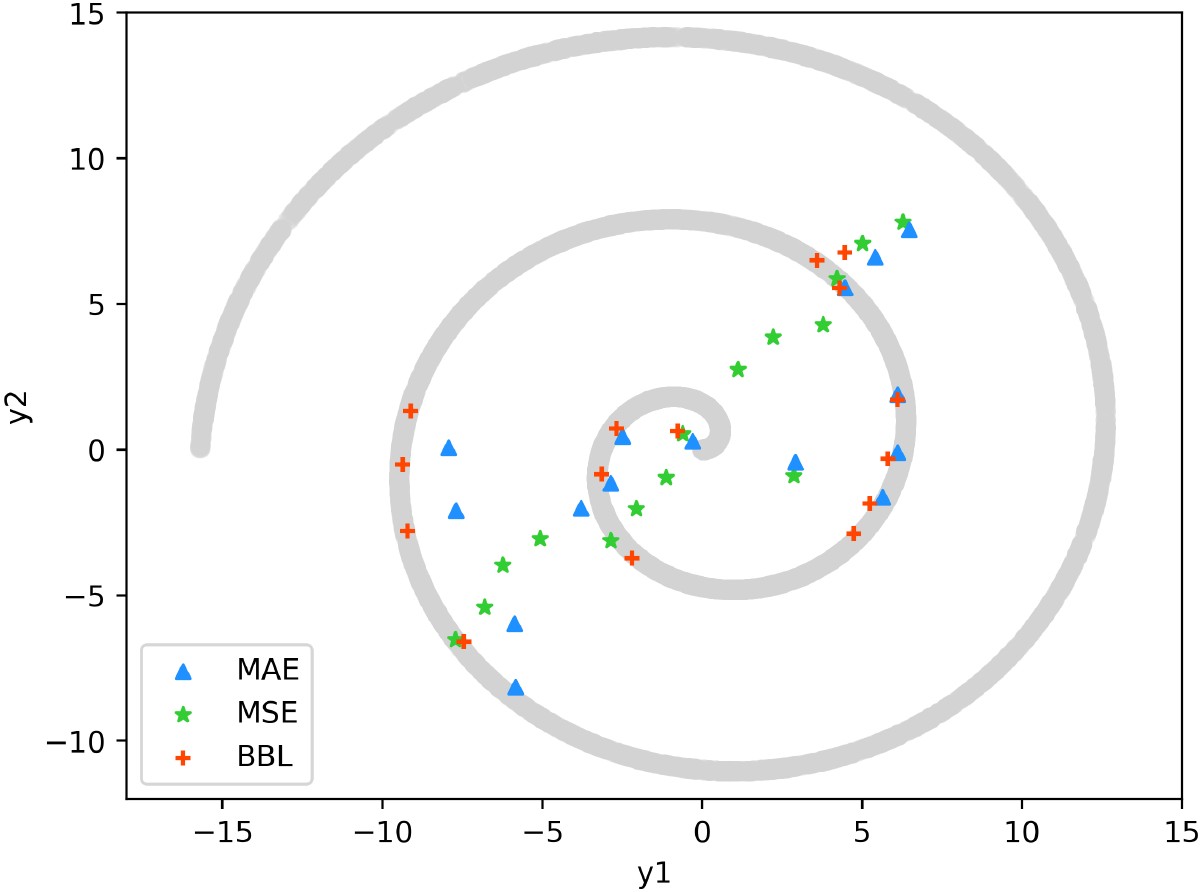}
	\end{center}
	\vspace{-0.15in}
	\caption{Toy example of \textbf{one-to-one} MAE, MSE and our \textbf{one-to-many} best-buddy loss (BBL) in SR. In training, two-dimensional HR data $y=\left[y_{1},y_{2}\right]$ is drawn from a Swiss-roll distribution (in gray). One-dimensional LR data is obtained by linearly downsampling HR data as $x=\frac{y_{1}+y_{2}}{2}$ like~\cite{sonderby2016amortised}.  In testing, we plot the HR estimates of MAE ($\begingroup\color{dodgerblue}\blacktriangle\endgroup$), MSE ($\begingroup\color{limegreen}\bigstar\endgroup$) and BBL ($\begingroup\color{red}\bm{+}\endgroup$) for $x \in \left[-7, -6, \cdots, 7\right]$. Our BBL fits the distribution best because it produces more HR estimates that fall on the roll.}
	\vspace{-0.15in}
	\label{fig:loss}
\end{figure}

It is well-known that SISR is essentially an ill-posed problem since a single low-resolution (LR) patch may correspond to multiple high-resolution (HR) solutions - it is difficult to decide the best. The commonly used one-to-one MSE/MAE loss tends to enforce a rigid mapping between the given LR-HR pair, which strictly constrains the HR space and makes it difficult to train the network. Relaxing the one-to-one constraint, we propose a novel {\it best-buddy loss}, an improved one-to-many MAE loss, to allow finding and using HR supervision signals flexibly by exploiting the ubiquitous self-similarity existing in natural images. Put it differently, an estimated HR patch is allowed to be supervised by different but close-to-ground-truth (GT) patches sourced from multiple scales of the corresponding GT image. From the toy example in Figure~\ref{fig:loss}, it is noticeable that our best-buddy loss outperforms one-to-one MSE/MAE. Additionally, a back-projection constraint is adopted to ensure the validity of the estimated HR signal.

As aforementioned, undesirable artifacts may be produced in images for existing GAN-based methods~\cite{johnson2016perceptual,ledig2017photo,mechrez2018maintaining,wang2018esrgan,zhang2019ranksrgan,soh2019natural}. We propose a region-aware adversarial learning strategy to address it. Our network treats smooth and well-textured areas differently, and only performs the adversarial training on the latter. This separation encourages the network to focus more on regions with rich details while avoiding generating unnecessary texture on flat regions (e.g., sky and building). With this improvement, our proposed best-buddy GANs (termed as {\it Beby-GAN}) is able to reconstruct photo-realistic high-frequencies with fewer undesirable artifacts (see Figure~\ref{fig:teasing}).

In summary, our contribution is threefold:
\begin{itemize}
	\item We present Beby-GAN for high-quality image super-resolution. The proposed \textit{one-to-many} best-buddy loss benefits generating richer and more plausible texture. Extensive experiments and a user study justify the effectiveness of the proposed method.
	\item A region-aware adversarial learning strategy is designed to further enhance the visual quality of images.
	\item Breaking through the 2K resolution limitation of current SISR datasets, we provide an ultra-high-resolution 4K (UH4K) image dataset with diverse categories to promote future research, which will be made publicly available. 
\end{itemize}

\section{Related Work}
Single image super-resolution (SISR) is a classical image restoration task. It is roughly divided into three categories of example-based or prior-based methods~\cite{yang2008image,glasner2009super,yang2010exploiting,zeyde2010single,timofte2013anchored,romano2014single,timofte2014a+}, PSNR-oriented methods~\cite{dong2014learning,kim2016accurate,kim2016deeply,shi2016real,lai2017deep,tai2017image,tai2017memnet,lim2017enhanced,haris2018deep,zhang2018residual,zhang2018image} and perception-driven methods~\cite{johnson2016perceptual,ledig2017photo,sajjadi2017enhancenet,mechrez2018maintaining,wang2018esrgan,wang2018fully,wang2018recovering,zhang2019ranksrgan,fritsche2019frequency,soh2019natural}.

\subsection{Example-Based Methods}
This line~\cite{yang2008image,zeyde2010single,timofte2014a+,yang2012coupled,peleg2014statistical} learns mapping from low-resolution patches to high-resolution counterparts, where the paired patches are collected from an external database. In this paper, we exploit this idea to search for one-to-many LR-HR mappings to produce visually pleasing results.

\begin{figure*}[t]
	\begin{center}
		\includegraphics[width=1.0\linewidth]{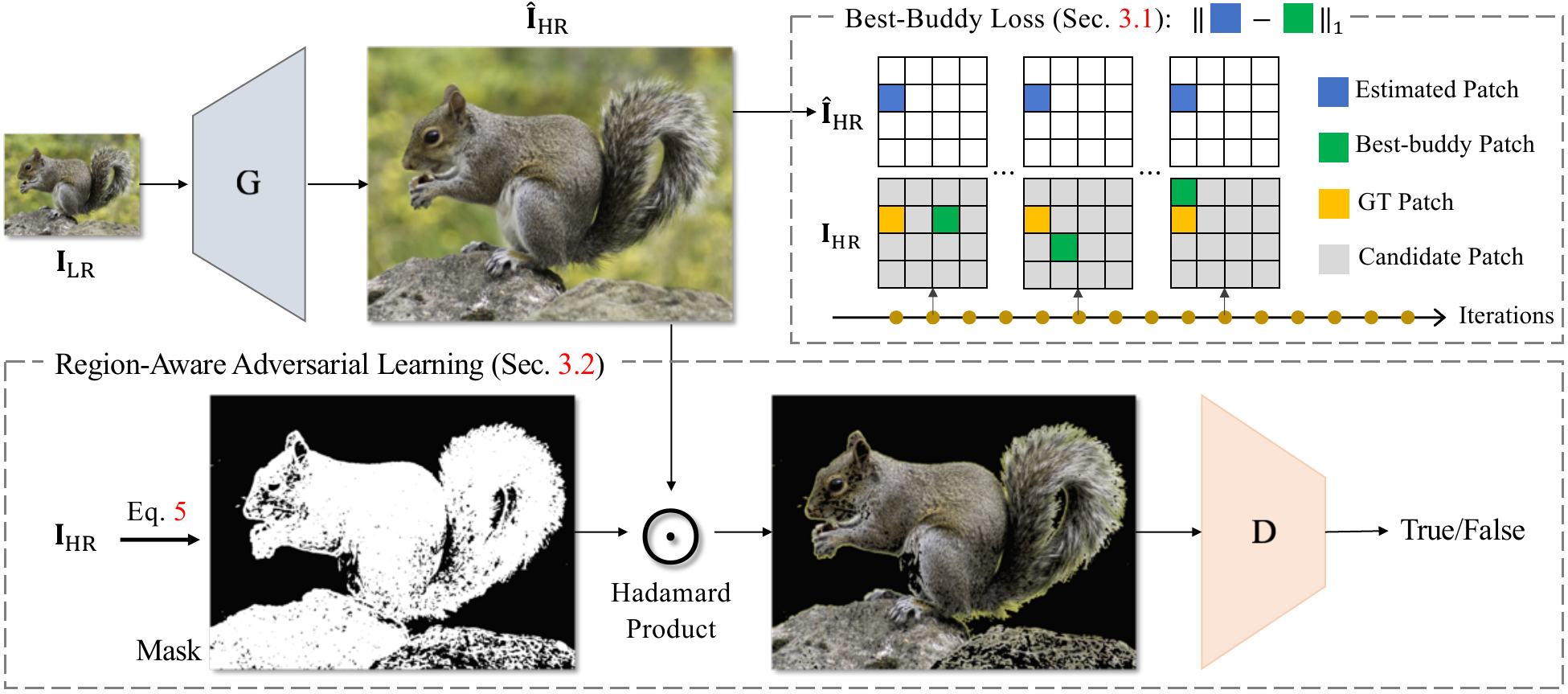}
	\end{center}
	\vspace{-0.15in}
	\caption{The framework of our proposed Beby-GAN. The best-buddy loss allows the estimated HR patches to be \textit{optimally} supervised in a dynamic way during training, as detailed in Sec.~\ref{sec:bbl} and Fig.~\ref{fig:bbl}(c). The region-aware adversarial learning is proposed to make the discriminator focus on rich-texture areas.}
	\vspace{-0.1in}
	\label{fig:framework}
\end{figure*}

\subsection{PSNR-Oriented Methods}
In past years, particular attention is paid to improve the pixel-wise reconstruction measures (e.g., peak-to-noise ratio, PSNR). It is the first time that SRCNN~\cite{dong2014learning} introduces a deep convolutional neural network into the SISR task. Afterwards, more well-designed architectures were proposed including residual and recursive learning~\cite{kim2016accurate,kim2016deeply,tai2017image,lim2017enhanced}, sub-pixel upsampling~\cite{shi2016real}, Laplacian pyramid structure~\cite{lai2017deep} and dense connecting~\cite{zhang2018residual}. Especially, Zhang \etal~\cite{zhang2018image} integrated channel attention modules into a network achieving a significant improvement in terms of PSNR performance. Hu \etal~\cite{hu2019meta} proposed Meta-SR to enable a single network to upsample an image into arbitrary sizes. Hussein \etal~\cite{hussein2020correction} designed a closed-form correction filter to convert an unknownly degraded image to a counterpart obtained from a standard bicubic kernel, and thus improved the generalization ability of pretrained models. 

\subsection{Perception-Driven Methods}
Despite breakthroughs on PSNR, the aforementioned methods still face a challenge that super-resolved images are typically overly-smooth and lack high-frequencies. To tackle this problem, Johnson \etal~\cite{johnson2016perceptual} proposed a novel perceptual loss. Ledig \etal ~\cite{ledig2017photo} presented SRGAN, which utilizes an adversarial loss and a content loss to push outputs into residing on the manifold of natural images. Thanks to a patch-wise texture loss, EnhanceNet~\cite{sajjadi2017enhancenet} obtains better performance on both quantitative and qualitative evaluation. ESRGAN~\cite{wang2018esrgan} marked a new milestone, which consistently generates more realistic texture benefiting from model and loss improvements. Later on, Zhang \etal~\cite{zhang2019ranksrgan} proposed RankSRGAN capable of being optimized towards a specific perceptual metric. In order to avoid undesirable artifacts, Soh \etal~\cite{soh2019natural} introduced a naturalness prior to constrain the reconstructed image in the natural manifold. However, most of these methods rely on single-LR-single-HR MSE/MAE supervision. Besides, without a region-aware mechanism, the architecture design can not deal with regions differently according to their properties. From these perspectives, we propose best-buddy GANs detailed as follows.

\section{Beby-GAN for Image Super-Resolution}
Given a low-resolution (LR) image $\mathbf{I}_{\rm LR} \in \mathbb{R}^{H \times W}$, single image super-resolution (SISR) is supposed to generate a high-resolution (HR) counterpart $\mathbf{\hat I}_{\rm HR} \in \mathbb{R}^{Hs \times Ws}$ under an upscale factor $s$. As shown in Figure~\ref{fig:framework}, the main body of our framework is built upon the generative adversarial networks (GANs)~\cite{goodfellow2014generative}, where the generator is used to reconstruct high-resolution images and the discriminator is trained to distinguish between recovered results and real natural images. Following~\cite{wang2018esrgan}, we adopt a pretrained RRDB model as our generator since it has demonstrated strong learning ability. In this section, we first describe the proposed best-buddy loss and region-aware adversarial learning strategy, followed by explaining other loss functions.

\subsection{Best-Buddy Loss}
\label{sec:bbl}
\begin{figure}[t]
	\begin{center}
		\includegraphics[width=1.0\linewidth]{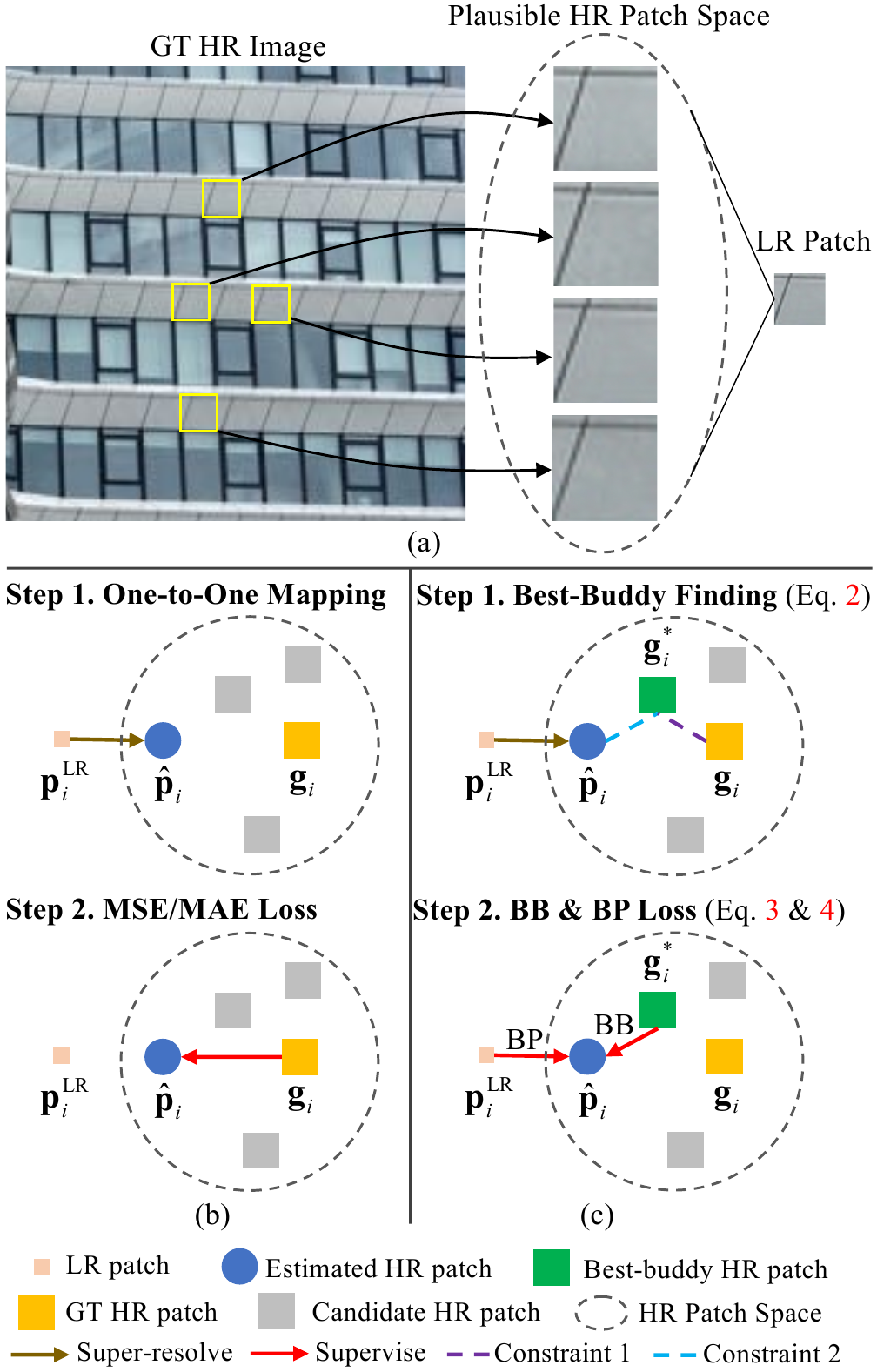}
	\end{center}
	\vspace{-0.1in}
	\caption{Comparison of the MSE/MAE and the best-buddy (BB) loss with a back-projection (BP) constraint. (a) Depiction of one-to-many nature in the SISR task. (b) MSE/MAE loss. (c) BB and BP loss. Variables $\mathbf{p}_{i}^{\rm LR}$, $\mathbf{\hat p}_{i}$, $\mathbf{g}_{i}$ and $\mathbf{g}_{i}^{*}$ indicate the LR patch, estimated HR patch, ground-truth HR patch and best-buddy HR patch in current iteration. } 
	\vspace{-0.1in}
	\label{fig:bbl}
\end{figure}

In the super-resolution task, a single LR patch is essentially associated with multiple natural HR solutions, as illustrated in Figure~\ref{fig:bbl}(a). Existing methods generally focus on learning immutable single-LR-single-HR mapping using an MSE/MAE loss in the training phase (see Figure~\ref{fig:bbl}(b)), which ignores the inherent uncertainty of SISR. As a result, the reconstructed HR images may lack several high-frequency structures~\cite{mathieu2016deep,johnson2016perceptual,dosovitskiy2016generating,estrach2016super,ledig2017photo}. 

To alleviate this issue, we propose a one-to-many best-buddy loss to enable trustworthy but much more flexible supervision. The key idea is that an estimated HR patch is allowed to be supervised by diverse targets in different iterations (see Figure~\ref{fig:framework}). In this paper, all supervision candidates come from a multi-scale ground-truth image pyramid. As shown in Figure~\ref{fig:bbl}(c), for an estimated HR patch $\mathbf{\hat p}_{i}$, we look for its corresponding supervision patch $\mathbf{g}_{i}^{*}$ (i.e., {\it best buddy}) in the current iteration to meet two constraints:

\vspace{0.05in}
\noindent{\textbf{Constraint 1.}} $\mathbf{g}_{i}^{*}$ is required to be close to the predefined ground-truth $\mathbf{g}_{i}$ in the HR space ($1^{st}$ term in Eq.~\ref{eq:bbf}). Relying on the ubiquitous multi-scale self-similarity in natural images~\cite{kindermann2005deblurring,protter2008generalizing,glasner2009super,yang2010exploiting,huang2015single,li2020mucan}, it is very likely to find a HR patch that is close to the ground-truth $\mathbf{g}_{i}$. 

\vspace{0.05in}
\noindent{\textbf{Constraint 2.}} In order to make optimization easy, $\mathbf{g}_{i}^{*}$ ought to be close to the estimation $\mathbf{\hat p}_{i}$ ($2^{nd}$ term in Eq.~\ref{eq:bbf}).  Note that $\mathbf{\hat p}_{i}$ is considered to be a reasonable prediction as our generator is well initialized. 
\vspace{0.05in}

Optimized with these two objectives, the obtained best buddy $\mathbf{g}_{i}^{*}$ is regarded as a plausible HR target for supervision. In detail, we first downsample the ground-truth (GT) HR image $\mathbf{I}_{\rm HR}$ with different scale factors as
\begin{equation}
	\begin{aligned}
		\mathbf{I}_{\rm HR}^{r} = S \left( \mathbf{I}_{\rm HR}, r \right) , \; r=\{2,4\} \,,
	\end{aligned}
\end{equation}
where $S(\mathbf{I}, r): \mathbb{R}^{H \times W} \rightarrow \mathbb{R}^{\frac{H}{r} \times \frac{W}{r}}$ is a bicubic downsampling operator, and obtain a 3-level image pyramid (including the original GT HR image). Then, we unfold the estimated HR image and corresponding GT image pyramid into patches ($3 \times 3$ in our paper), among which the GT part forms the supervision candidate database $\mathcal{G}$ of this image. For the $i$-th estimated patch $\mathbf{\hat p}_{i}$, instead of being supervised by the immutable predefined GT patch $\mathbf{g}_{i}$, it is allowed to find the best buddy $\mathbf{g}_{i}^{*}$ in the current iteration as
\begin{equation}
	\begin{aligned}
		\mathbf{g}_{i}^{*} = \underset{\mathbf{g} \in \mathcal{G}}{\arg\min} \ \alpha {\left\| \mathbf{g} - \mathbf{g}_{i}\right\|}_{2}^{2} + \beta {\left\| \mathbf{g} - \mathbf{\hat p}_{i} \right\|}_{2}^{2}\,,
	\end{aligned}
	\label{eq:bbf}
\end{equation}
where  $\alpha \ge 0$ and $\beta  \ge 0$ are scaling parameters. The best-buddy loss for this patch pair $\left( \mathbf{\hat p}_{i}, {\mathbf{g}}_{i}^{*} \right)$ is represented as
\begin{equation}
	\begin{aligned}
		\mathcal{L}_{\rm BB} \left( \mathbf{\hat p}_{i}, {\mathbf{g}}_{i}^{*}  \right) = \left\| \mathbf{\hat p}_{i} - {\mathbf{g}}_{i}^{*} \right\|_{1} \,.
	\end{aligned}
\end{equation}
Notice that when $\beta \ll \alpha$, the proposed best-buddy loss degenerates into the typical one-to-one MAE loss.

\begin{figure}[t]
	\begin{center}
		\includegraphics[width=1.0\linewidth]{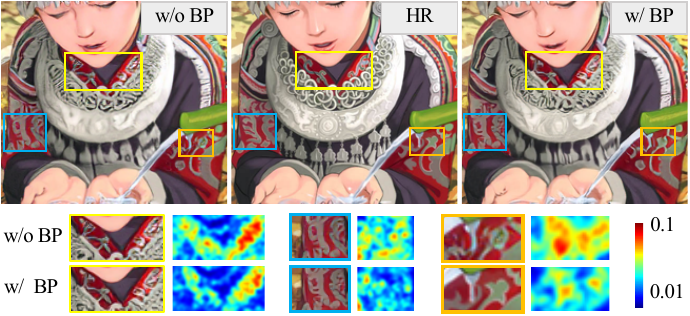}
	\end{center}
	\vspace{-0.15in}
	\caption{Comparison between with (w/) and without (w/o) the back-projection (BP) loss. We visualize the L2 error heatmaps between the estimated results and ground truth.}
	\vspace{-0.15in}
	\label{fig:bp}
\end{figure}

Besides, we enforce another back-projection constraint on the estimation $\mathbf{\hat I}_{\rm HR}$. The super-resolved images after downscaling must match the fidelity expected at the lower resolution. We introduce an HR-to-LR operation (bicubic downsampling in this paper) to ensure that the projection of the estimated HR image onto the LR space is still consistent with the original LR one. The back-projected result is supervised by
\begin{equation}
	\begin{aligned}
		\mathcal{L}_{\rm BP} = \left\| S \left( \mathbf{\hat I}_{\rm HR},s \right) - \mathbf{I}_{\rm LR} \right\|_{1} \,,
	\end{aligned}
\end{equation}
where $s$ is the downscale factor. From Figure~\ref{fig:bp}, we notice that this back-projection loss plays an essential role in maintaining content and color consistency.

\subsection{Region-Aware Adversarial Learning}
\label{sec:ra}
As shown in Figure~\ref{fig:teasing}, previous GAN-based methods sometimes produce undesirable texture, especially in flat regions. Thus, as illustrated in Figure~\ref{fig:framework}, we propose to differentiate the rich-texture areas from smooth ones according to local pixel statistics, and only feed the textured content to the discriminator since smooth regions can be easily recovered without adversarial training. Our strategy is to first unfold the ground-truth HR image (i.e., $\mathbf{I}_{\rm HR} \in \mathbb{R}^{Hs \times Ws}$ ) into patches $\mathbf{B} \in \mathbb{R}^{Hs \times Ws \times k^{2}}$ with size $k^{2}$,  and then compute standard deviation (std) for each patch. After that, a binary mask is obtained as
\begin{equation}
	\begin{aligned}
		\textbf{M}_{i,j} &=
		\begin{cases}
			1, &  {\rm std} \left( \mathbf{B}_{i,j} \right) \ge \delta \\
			0, &  {\rm std} \left( \mathbf{B}_{i,j} \right) < \delta \,,
		\end{cases}
	\end{aligned}
\end{equation}
where $\delta$ is a predefined threshold and $(i, j)$ is the patch location. The highly textured regions are marked as 1 while flat regions as 0. Then the estimated result $\mathbf{\hat I}_{\rm HR}$ and ground-truth $\mathbf{I}_{\rm HR}$ are multiplied with the same mask $\mathbf{M}$ yielding $\mathbf{\hat I}_{\rm HR}^{\rm \mathbf{M}}$ and $\mathbf{I}_{\rm HR}^{\rm \mathbf{M}}$, which are next processed by the following discriminator. Though more sophisticated strategies can be used at the cost of more computation, we show that this simple masking method already works very well in Sec.~\ref{sec:ablation}. It directs our model to generate realistic fine details for textured areas.

\subsection{Other Loss Functions}

\noindent\textbf{Perceptual Loss.} Apart from computing pixel-wise distances in image space, several works~\cite{estrach2016super,dosovitskiy2016generating,johnson2016perceptual} use feature similarity to enhance the perceptual quality of recovered images. Following this idea, we set the perceptual loss as
\begin{equation}
	\begin{aligned}
		\mathcal{L}_{\rm P} = \sum_{i}{ \eta_{i} \left\| \phi_{i} \left( \mathbf{\hat I}_{\rm HR} \right) - \phi_{i} \left (\mathbf{I}_{\rm HR} \right) \right\|_{1} } \,,
	\end{aligned}
\end{equation}
where $\phi_{i}$ represents the $i$-th layer activation of a pretrained VGG-19~\cite{Simonyan15} network and $\eta_{i}$ is a scaling coefficient. To capture feature representations at different levels, we take into consideration three layers including $conv_{3\_4}, conv_{4\_4}$ and $conv_{5\_4}$ and set scaling coefficients to $\frac{1}{8}, \frac{1}{4}$ and $\frac{1}{2}$.

\vspace{0.05in}
\noindent\textbf{Adversarial Loss.} The discriminator in our network is implemented based on Relativistic average GANs (RaGANs)~\cite{jolicoeur2018relativistic}, which estimate the probability that a ground-truth HR image is more realistic than a generated one. It has been shown that RaGANs are more stable and can produce results of higher quality~\cite{jolicoeur2018relativistic,wang2018esrgan,soh2019natural}. The loss functions are formulated as
\begin{equation}
	\begin{aligned}
		\mathcal{L}_{\rm D} = - \mathbb{E}_{x_{r} \sim \mathbb{P}} \left[ \log \left( \bar{D} \left( x_{r} \right) \right) \right] - \mathbb{E}_{x_{f} \sim \mathbb{Q}} \left[ \log \left( 1 - \bar{D} \left( x_{f} \right) \right) \right] \,,
	\end{aligned}
\end{equation}
\begin{equation}
	\begin{aligned}
		\mathcal{L}_{\rm G} = - \mathbb{E}_{x_{r} \sim \mathbb{P}} \left[ \log \left(1 - \bar{D} \left( x_{r} \right) \right) \right] - \mathbb{E}_{x_{f} \sim \mathbb{Q}} \left[ \log \left(\bar{D} \left( x_{f} \right) \right) \right] \,,
	\end{aligned}
\end{equation}
where
\begin{equation}
	\begin{aligned}
		\bar{D} \left( x \right) &=
		\begin{cases}
			{\rm sigmoid} \left( C \left( x \right) - \mathbb{E}_{x_{f} \sim \mathbb{Q}} C \left( x_{f} \right) \right), & {\rm x \ is \ real} \\
			{\rm sigmoid} \left( C \left( x \right) - \mathbb{E}_{x_{r} \sim \mathbb{P}} C \left( x_{r} \right) \right), & {\rm x \ is \ fake} \,. \\
		\end{cases}
	\end{aligned}
	\label{eq:9}
\end{equation}
In Eq.~\ref{eq:9}, $x_{r}$ denotes the masked real data sampled from distribution $\mathbb{P}$, $x_{f}$ is the masked fake data (i.e., generated data) sampled from distribution $\mathbb{Q}$ and $C\left( x \right)$ is the non-transformed discriminator output.

\vspace{0.05in}
\noindent\textbf{Overall Loss.} The overall loss of the generator is formulated as:
\begin{equation}
	\begin{aligned}
		\mathcal{L} = \lambda_{1}\mathcal{L}_{\rm BB} + \lambda_{2}\mathcal{L}_{\rm BP} + \lambda_{3}\mathcal{L}_{\rm P} + \lambda_{4}\mathcal{L}_{\rm G} \,,
	\end{aligned}
\end{equation}
where $\lambda_{1}=0.1$, $\lambda_{2}=1.0$, $\lambda_{3}=1.0$ and $\lambda_{4}=0.005$.

\begin{figure}[t]
	\begin{center}
		\includegraphics[width=1.0\linewidth]{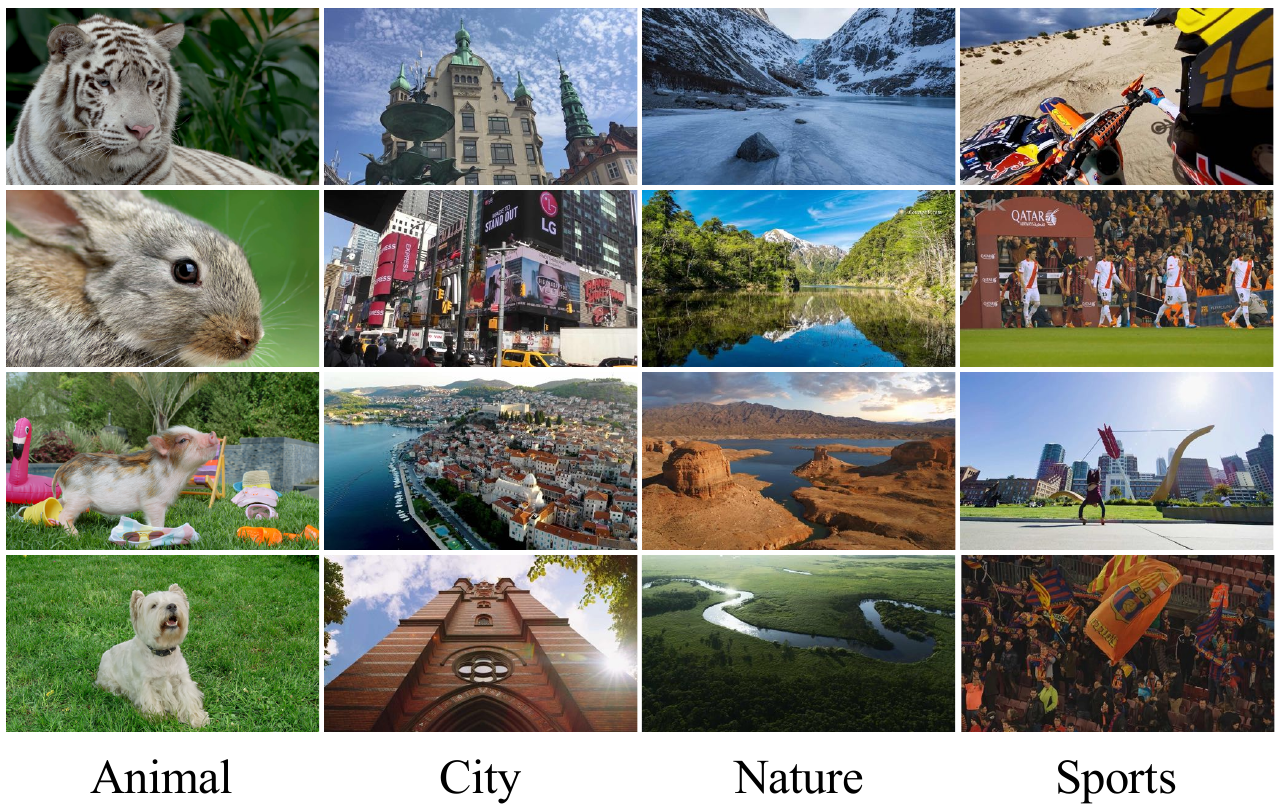}
	\end{center}
	\vspace{-0.2in}
	\caption{Visualization of examples from four categories in our proposed UH4K dataset.}
	\vspace{-0.1in}
	\label{fig:thumbnail}
\end{figure}

\section{Experiments}

\subsection{Datasets and Metrics}
Following previous methods~\cite{wang2018esrgan,zhang2019ranksrgan}, our network is trained on DIV2K~\cite{agustsson2017ntire} (800 images) and Flickr2K~\cite{timofte2017ntire} (2650 images) datasets. Apart from the widely used testing benchmark including Set5~\cite{bevilacqua2012low}, Set14~\cite{zeyde2010single}, BSDS100~\cite{martin2001database} and Urban100~\cite{huang2015single}, we also adopt the 100 validation images in DIV2K to evaluate the performance of our model.

Besides, we propose an ultra-high-resolution 4K (UH4K) dataset to perform a more challenging and complete study on the single image super-resolution (SISR) task. The images are collected from YouTube with resolution $3840 \times 2160$. The dataset contains over 400 images featuring four categories, i.e., animal, city, nature and sports. A few examples are visualized in Figure~\ref{fig:thumbnail}. Compared with existing benchmark datasets, ours has higher resolution, higher variety and richer texture/structure. In this paper, our 4K dataset is \textit{only} used for evaluation.

Apart from PSNR and SSIM~\cite{wang2004image} (the higher, the better ideally), we adopt a ground-truth-based perceptual metric LPIPS~\cite{zhang2018unreasonable} and a no-reference perceptual metric PI~\cite{blau20182018} (the lower, the better ideally) in our experiments. Besides, we conduct a user study to make a better comparison. 

\subsection{Training Details}
All experiments are carried out on NVIDIA GeForce RTX 2080 Ti GPUs under the $\times 4$ setting. The mini-batch size is set to 8. We adopt Adam as the optimizer with $\beta_{1}=0.9$ and $\beta_{2}=0.999$. There are 3 periods in our training, each with $200 {\rm K}$ iterations. The learning rate for every period is set to $1e^{-4}$ initially in accompany with a warm-up and a cosine decay. The images are augmented with random cropping, flipping and rotation. The input size is $48 \times 48$ and the rotation is $90^{\circ}$ or $-90^{\circ}$. The $\alpha$ and $\beta$ in Sec.~\ref{sec:bbl} are both set to 1.0. The kernel size $k$ and $\delta$ in Sec.~\ref{sec:ra} are set to 11 and 0.025 (for normalized images) from empirical experiments. The calculation of bset-buddy loss only costs 5.9ms for images of size $196 \!\times\! 196$.

\subsection{Ablation Study}
\label{sec:ablation}
\begin{figure}[t]
	\begin{center}
		\includegraphics[width=0.95\linewidth]{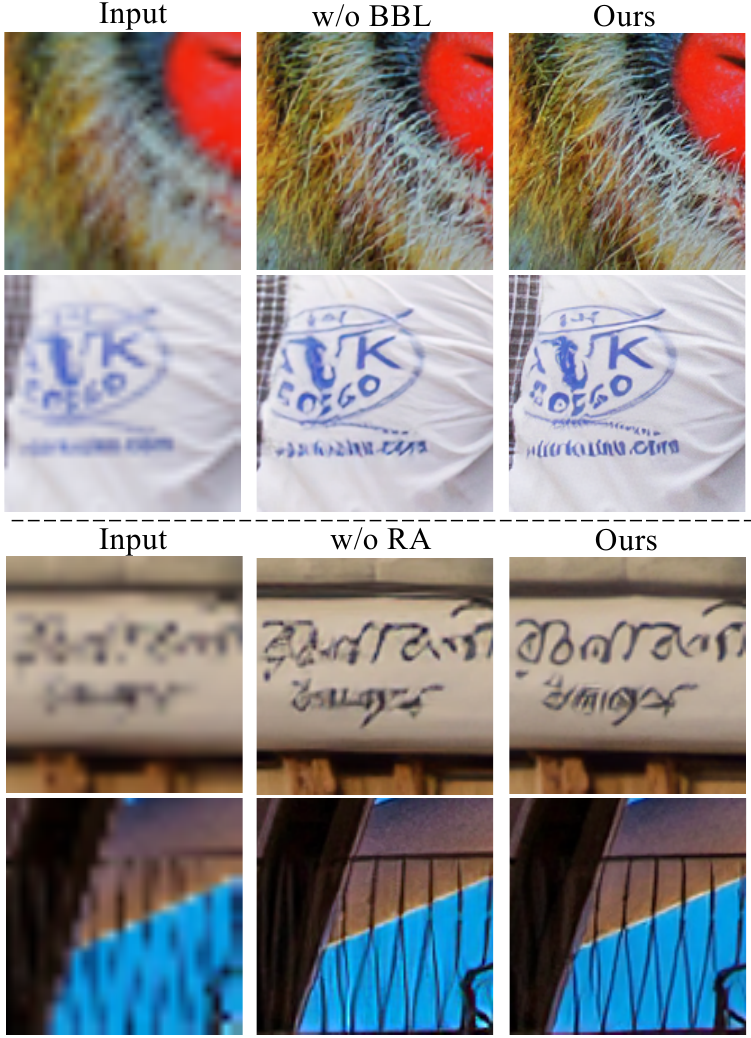}
	\end{center}
	\vspace{-0.15in}
	\caption{Visual comparison for ablation study.  ``Ours'' is our proposed Beby-GAN. ``w/o BBL'' and ``w/o RA'' means removing best-buddy loss and region-aware learning.}
	\vspace{-0.1in}
	\label{fig:abl}
\end{figure}

In this part, we investigate how each design affects the perceptual quality of super-resolved images. Starting from our best-buddy GANs (Beby-GAN), we ablate the best-buddy loss and region-aware learning strategy, respectively. We show a visual comparison in Figure~\ref{fig:abl}. Also, we evaluate the LPIPS performance  in Table~\ref{tab:abla} because the ground-truth-based LPIPS is more consistent with human perception compared with no-reference perceptual metrics illustrated in Sec.~\ref{sec:comparison}. All these results verify the effectiveness of our method.

\begin{table}[t]
	\renewcommand\arraystretch{1.1}
	\begin{center}
		\resizebox{0.42\textwidth}{!}{
			\begin{tabular}{ c | c | c | c }
				\hline
				Models & Ours & w/o BBL & w/o RA \\
				\hline
				Set14~\cite{zeyde2010single} & \textbf{0.2202}   &  0.2343 & 0.2272 \\
				BSDS100~\cite{martin2001database} &  \textbf{0.2385} & 0.2514 & 0.2469 \\
				\hline
		\end{tabular}}
	\end{center}
	\vspace{-0.15in}
	\caption{LPIPS comparison for ablation study. ``Ours'' is our proposed Beby-GAN. ``w/o BBL'' and ``w/o RA'' indicate removing best-buddy loss and region-aware learning.}
	\vspace{-0.12in}
	\label{tab:abla}
\end{table}

\vspace{0.05in}
\noindent\textbf{Best-buddy loss.} In contrast to the commonly used one-to-one MSE/MAE loss, our best-buddy loss allows the network to learn single-LR-multiple-HR mapping. As illustrated in Figure~\ref{fig:abl}, our BBL (see ``Ours'') recovers richer texture and sharper edges compared with one-to-one MAE (see `` w/o BBL''). The whiskers have more high-frequency details and the text is clearer. Also, we notice that the super-resolved images are more natural and visually pleasing. As shown in Table~\ref{tab:abla}, the best-buddy loss brings about considerable  improvement on LPIPS results.

\vspace{0.05in}
\noindent\textbf{Region-Aware Adversarial Learning.} As shown in the Figure~\ref{fig:abl}, there exist unpleasant artifacts near the characters and railing in the results without region-aware learning (see ``w/o RA''). After differentiating between rich-texture and flat areas, this problem is alleviated as shown in the $3^{nd}$ column (see ``Ours''). The separation allows the network to know ``where'' to conduct the adversarial training and yields two major advantages. On the one hand, it leads to easier training since the network only needs to focus on regions of high-frequency details. On the other hand, the network produces less unnatural texture. The results in Table~\ref{tab:abla} also demonstrate the effectiveness of this strategy.
\vspace{0.05in}

\begin{table*}[t]
	\renewcommand\arraystretch{1.3}
	\begin{center}
		\resizebox{\textwidth}{!}{
			\begin{tabular}{ c | c | c c c | c  c  c  c  c  c }
				\hline
				\hline
				\multirow{2}*{Dataset} & \multirow{2}*{Metric} & \multicolumn{3}{c|}{PSNR-Oriented Methods}  & \multicolumn{6}{c}{GAN-Based Methods}\\
				\cline{3-11}
				~ & ~ & SRResNet~\cite{ledig2017photo} & RRDB~\cite{wang2018esrgan} & RCAN~\cite{zhang2018image} & SRGAN~\cite{ledig2017photo} & ESRGAN-PI~\cite{wang2018esrgan} & ESRGAN~\cite{wang2018esrgan} & RankSRGAN~\cite{zhang2019ranksrgan}$^\dagger$  & \textbf{Beby-GAN-PI} & \textbf{Beby-GAN} \\
				\hline
				\multirow{4}*{Set14} & LPIPS$\downarrow$ & 0.3043 & 0.2934 & 0.2922 & 0.3162 & 0.2771 & 0.2372 & 0.2545 & 0.2537 & \textcolor{red}{0.2202} \\
				~ & PI$\downarrow$ & 5.36 & 5.27 & 5.30 & 2.87 & 2.61 & 2.93 & 2.61 & \textcolor{red}{2.57}  & 3.08 \\
				~ & PSNR$\uparrow$ & 28.57 & 28.95 & 28.97 & 25.90 & 26.39  & 26.28 & 26.57 & 26.56  & \textcolor{red}{26.96} \\
				~ & SSIM$\uparrow$ & 0.7834 & 0.7912 & 0.7913 & 0.6942 & 0.7021 & 0.6985 & 0.7052 & 0.7061 & \textcolor{red}{0.7282} \\
				\hline
				\multirow{4}*{BSDS100} & LPIPS$\downarrow$ & 0.3437 & 0.3341 & 0.3320 & 0.3387 & 0.2801 & 0.2599 & 0.2790 & 0.2777 & \textcolor{red}{0.2385} \\
				~ & PI$\downarrow$ & 5.34 & 5.30 & 5.19 & 2.62 & 2.27 & 2.48 & 2.15 & \textcolor{red}{2.13} & 2.44 \\
				~ & PSNR$\uparrow$ & 27.61 & 27.84 & 27.84 & 25.38 & 25.72 & 25.32 & 25.57 & 25.56 & \textcolor{red}{25.81} \\
				~ & SSIM$\uparrow$ & 0.7376 & 0.7453 & 0.7456 & 0.6423 & 0.6638 & 0.6514 & 0.6492 & 0.6536 & \textcolor{red}{0.6781} \\
				\hline
				\multirow{4}*{DIV2K} & LPIPS$\downarrow$ & 0.2991 & 0.2863 & 0.2862 & 0.3109 & 0.2741 & 0.2222 & 0.2368 & 0.2352 & \textcolor{red}{0.1991} \\
				~ & PI$\downarrow$ & 5.40 & 5.28 & 5.33 & 3.25 & \textcolor{red}{2.95} & 3.27 & 3.00 & 3.02 & 3.33 \\
				~ & PSNR$\uparrow$ & 30.49 & 30.90 & 30.86 & 27.16 & 27.80 & 28.16 & 28.01 & 28.12 & \textcolor{red}{28.71} \\
				~ & SSIM$\uparrow$ & 0.8391 & 0.8478 & 0.8469 &  0.7600 & 0.7653 & 0.7752 & 0.7652 & 0.7688 & \textcolor{red}{0.7923} \\
				\hline
				\multirow{4}*{UH4K} & LPIPS$\downarrow$ & 0.2304 & 0.2225 & 0.2208 & 0.3346 & 0.2694 & 0.2160 & 0.2745 & 0.2711 & \textcolor{red}{0.2009} \\
				~ & PI$\downarrow$ & 5.53 & 5.50 & 5.56 & 3.91 & 2.93 & 3.42 & \textcolor{red}{2.87} & 2.89 & 3.54 \\
				~ & PSNR$\uparrow$ & 32.11 & 32.45 & 32.46 & 27.85 & 28.94 & 29.43 & 28.99 & 29.13 & \textcolor{red}{30.02} \\
				~ & SSIM$\uparrow$ & 0.8691 & 0.8756 & 0.8751 & 0.7706 & 0.7874 & 0.8052 & 0.7850 & 0.7892 & \textcolor{red}{0.8214} \\
				\hline
		\end{tabular}}
	\end{center}
	\vspace{-0.15in}
	\caption{Quantitative comparison of LPIPS~\cite{zhang2018unreasonable}, PI~\cite{blau20182018}, PSNR and SSIM~\cite{wang2004image} on benchmarks. `$\downarrow$' means the lower, the better. `$\uparrow$' indicates the higher, the better. `$^\dagger$' means that the results of RankSRGAN~\cite{zhang2019ranksrgan} are from multiple models optimized by different objectives. \textcolor{red}{Red}: \textcolor{red}{best} results in GAN-based methods.}
	\vspace{-0.12in}
	\label{tab:quan}
\end{table*}

\subsection{Comparison with State-of-the-Art Methods}
\label{sec:comparison}
We compare our Beby-GAN with start-of-the-art methods of two categories. They are PSNR-oriented methods including SRResNet~\cite{ledig2017photo}, RRDB~\cite{wang2018esrgan}, RCAN~\cite{zhang2018image}, and perception-driven methods including ESRGAN~\cite{wang2018esrgan} and RankSRGAN~\cite{zhang2019ranksrgan}. We use Set14~\cite{zeyde2010single}, BSDS100~\cite{martin2001database}, DIV2K validation~\cite{agustsson2017ntire} and a subset of our UH4K for quantitative evaluation while more datasets~\cite{bevilacqua2012low,huang2015single,matsui2017sketch} for qualitative analysis.

\begin{figure}[t]
	\begin{center}
		\includegraphics[width=0.9\linewidth]{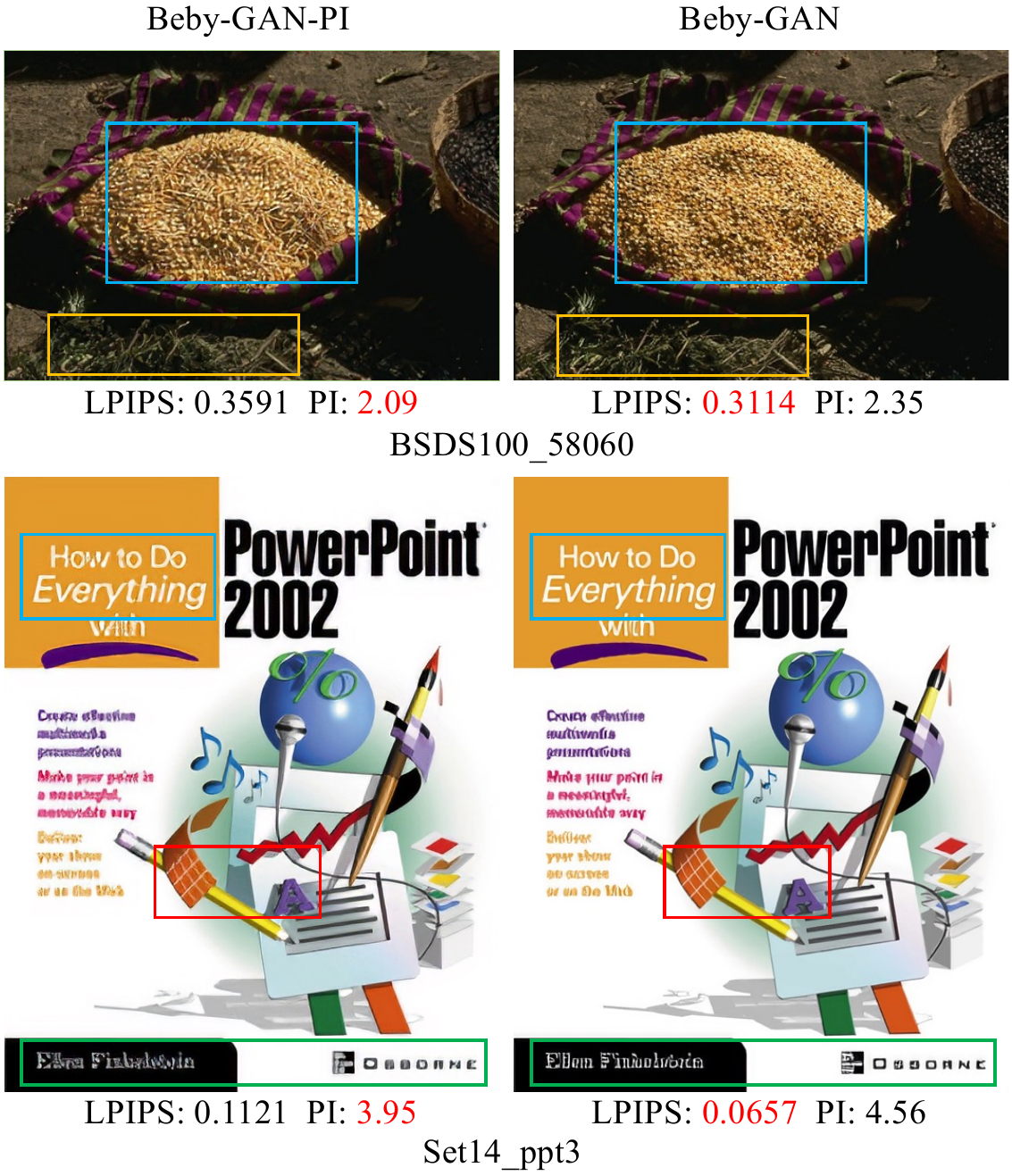}
	\end{center}
	\vspace{-0.15in}
	\caption{Comparison of Beby-GAN-PI and Beby-GAN. \textcolor{red}{Red}: \textcolor{red}{better} results. Ground-truth-based LPIPS is more representative and robust than no-reference-based PI. Zoom in for better visual comparison.}
	\vspace{-0.25in}
	\label{fig:pi}
\end{figure}

\vspace{-0.15in}
\subsubsection{Quantitative Results}
Since PSNR and SSIM~\cite{wang2004image} have already been shown to correlate weakly with the human perception regarding image quality, we further utilize LPIPS~\cite{zhang2018unreasonable} and PI~\cite{blau20182018} for evaluation. Following ESRGAN~\cite{wang2018esrgan} and RankSRGAN~\cite{zhang2019ranksrgan} that provide different models for quantitative evaluation, we prepare two models, named Beby-GAN-PI and Beby-GAN. The former is obtained using similar approaches as ESRGAN-PI~\cite{wang2018esrgan}. As shown in Table~\ref{tab:quan}, GAN-based methods obtain better performance on perceptual metrics with lower PSNR/SSIM scores.

As for GAN-based methods, our Beby-GAN performs best on PSNR/SSIM measures. Also, our method yields new state-of-the-art in terms of LPIPS on all benchmark. In terms of PI, our PI-based model achieves superior performance on Set14~\cite{zeyde2010single} and BSDS100~\cite{martin2001database}, as well as comparable results on DIV2K~\cite{agustsson2017ntire} and UH4K. We notice that there is a relatively large disparity between the ground-truth-based LPIPS and no-reference-based PI. As shown in Figure~\ref{fig:pi}, LPIPS is more consistent with human perception. In this case, PI is only used for reference.


\vspace{-0.15in}
\subsubsection{Qualitative Results}
As illustrated in Figure~\ref{fig:visual}, PSNR-oriented methods (i.e., RRDB~\cite{wang2018esrgan} and RCAN~\cite{zhang2018image}) tend to generate overly-smooth results. Although existing GAN-based methods (i.e., ESRGAN~\cite{wang2018esrgan} and RankSRGAN~\cite{zhang2019ranksrgan}) can recover some details, they possibly generate unpleasing visual artifacts (see UH4K$\_$020, DIV2K$\_$0823 and Set14$\_$Flowers) and color inconsistency (see Set14$\_$Comic).

In contrast, our Beby-GAN is capable of producing more realistic results. From visualized examples in Figure~\ref{fig:visual}, it is clear that our method reconstructs richer and more compelling patterns as well as sharper structures. Also, fewer artifacts are produced. In the following, we further present a comprehensive user study to evaluate the human visual quality of reconstructed images.  

\begin{figure*}[t]
	\begin{center}
		\includegraphics[width=1.0\linewidth]{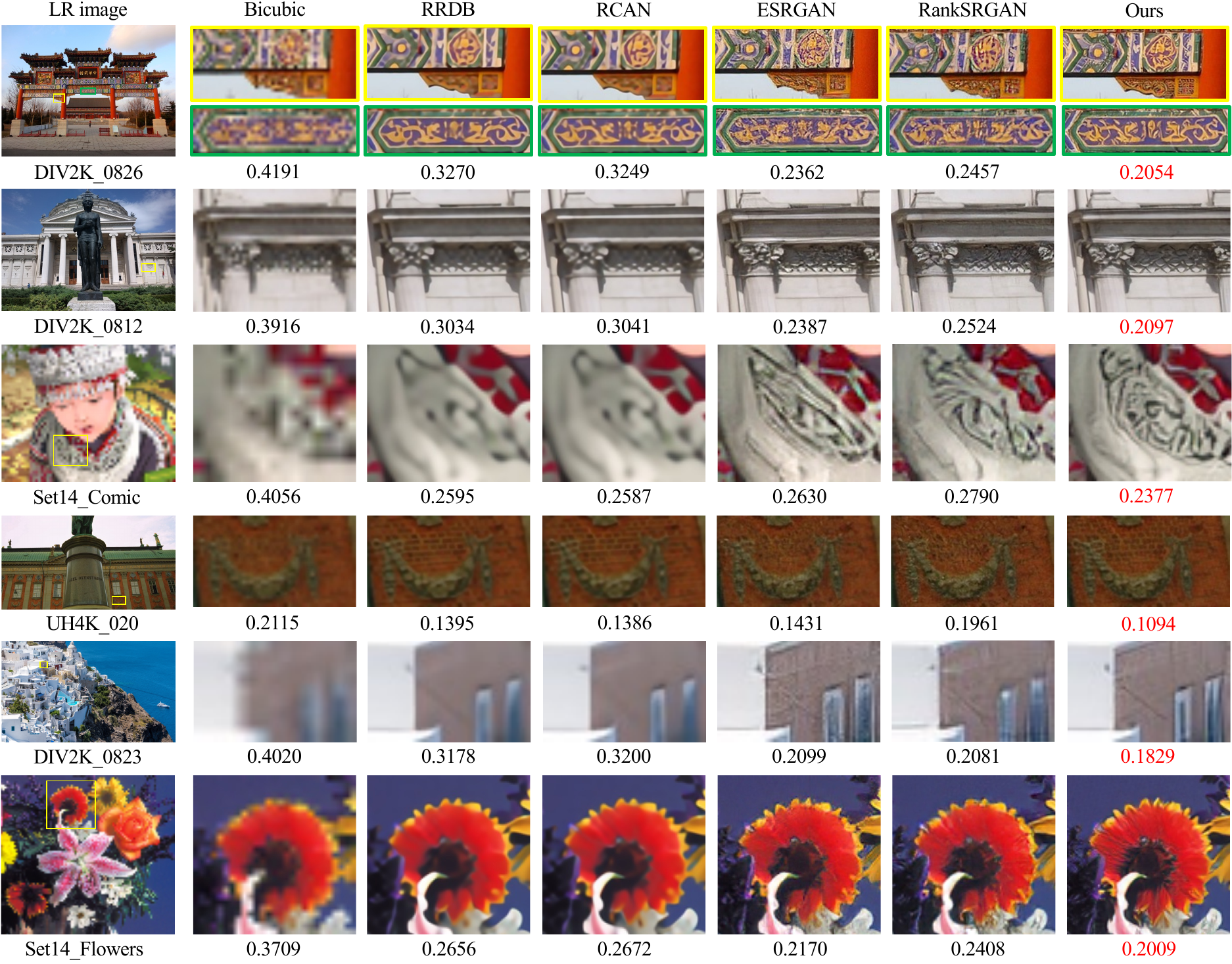}
	\end{center}
	\vspace{-0.15in}
	\caption{Visual comparison of our Beby-GAN with other methods on $\times 4$ scale. The values beneath images represent LPIPS measures. \textcolor{red}{Red}: best quantitative results. It is clear that our Beby-GAN obtains the best visual performance.}
	\vspace{-0.1in}
	\label{fig:visual}
\end{figure*}

\begin{figure}[t]
	\begin{center}
		\includegraphics[width=1.0\linewidth]{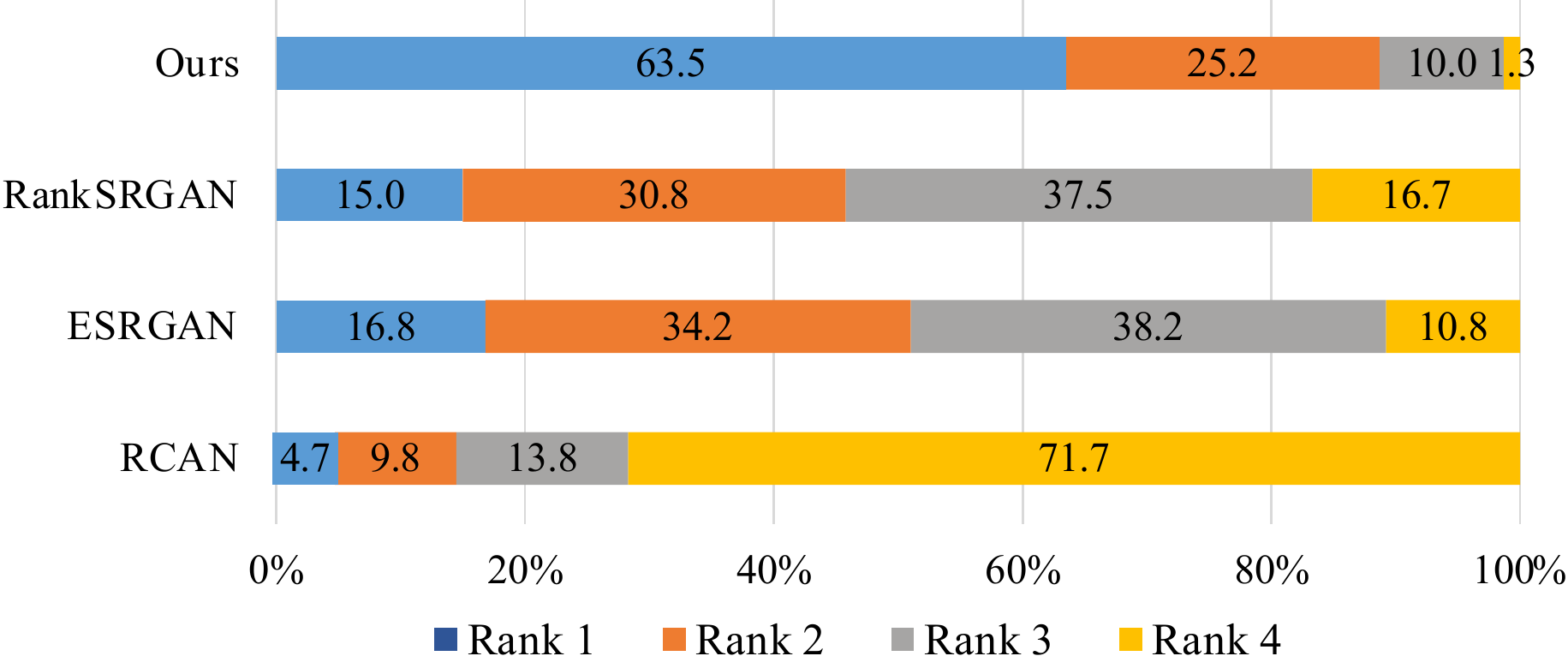}
	\end{center}
	\vspace{-0.12in}
	\caption{The ranking results of user study involving 30 participants. The values on the bars represent the percentages of rankings.}
	\vspace{-0.15in}
	\label{fig:user}
\end{figure}

\subsection{User Study}
\label{sec:user}
In addition to our method, we take into consideration RCAN~\cite{zhang2018image}, ESRGAN~\cite{wang2018esrgan} and RankSRGAN~\cite{zhang2019ranksrgan}. We prepare testing cases from three sources:
(1) \textit{Low-resolution images} stemming from the commonly used benchmark including Set5~\cite{bevilacqua2012low}, Set14~\cite{zeyde2010single}, BSDS100~\cite{martin2001database} and Urban100~\cite{huang2015single}. There are a total of 219 images.
(2) \textit{2K resolution} images from the validation subset of DIV2K~\cite{agustsson2017ntire}. 100 images are included.
(3) \textit{4K resolution} images in our UH4K dataset. There are over 400 images from 4 categories.

Every time we randomly display 30 testing cases and ask the participant to rank 4 versions of each image: RCAN, ESRGAN, RankSRGAN and our Beby-GAN. To make a fair comparison, we follow~\cite{zhang2019ranksrgan} to zoom in one small patch for each image.

We invite 30 participants to our user study. As shown in the Figure~\ref{fig:user}, most of our results rank in the first place while the remaining of ours still gets high rankings. Especially, it is observed that our Beby-GAN outperforms others by a large margin on the 4K dataset from visual examples in the appendix, generating more realistic details for this most challenging dataset. Besides, ESRGAN and RankSRGAN achieve better performance than RCAN. The user study not only demonstrates the superiority of our proposed Beby-GAN, but also explains that existing evaluation measures and human perception are diverse to some extent.

\section{Conclusion}
In this paper, we have presented best-buddy GANs (Beby-GAN) for highly detailed image super-resolution. By virtue of the proposed best-buddy loss and region-aware adversarial learning, our Beby-GAN is able to recover realistic texture while maintaining the naturalness of images. Extensive experiments along with the user study manifest the effectiveness of our method for SISR.


\clearpage

\setcounter{figure}{0}
\setcounter{section}{0}
\setcounter{table}{0}

\newpage
\twocolumn[
\begin{@twocolumnfalse}
\begin{center}
	{
		\Large
		\bf
		\begin{tabular}[t]{c}
			Best-Buddy GANs for Highly Detailed Image Super-Resolution \\ Appendix
		\end{tabular}
		\par
	}
	\vspace*{24pt}
	{
		\large
		\lineskip .5em
		\begin{tabular}[t]{c}
			Wenbo Li\textsuperscript{1}\printfnsymbol{1} \quad Kun Zhou\textsuperscript{2}\printfnsymbol{1} \quad Lu Qi\textsuperscript{1} \quad Liying Lu\textsuperscript{1} \quad Nianjuan Jiang\textsuperscript{2} \quad Jiangbo Lu\textsuperscript{2}\printfnsymbol{2} \quad Jiaya Jia\textsuperscript{1,2} \\
			$^{1}$The Chinese University of Hong Kong \quad ${^2}$Smartmore Corporation\\
			\vspace*{1pt}\\
		\end{tabular}
		\par
	}
	\vskip .5em
	\vspace*{8pt}
\end{center}
\end{@twocolumnfalse}
]




\section{Additional Qualitative Results}

\begin{figure*}[t]
	\begin{center}
		\includegraphics[width=1.0\linewidth]{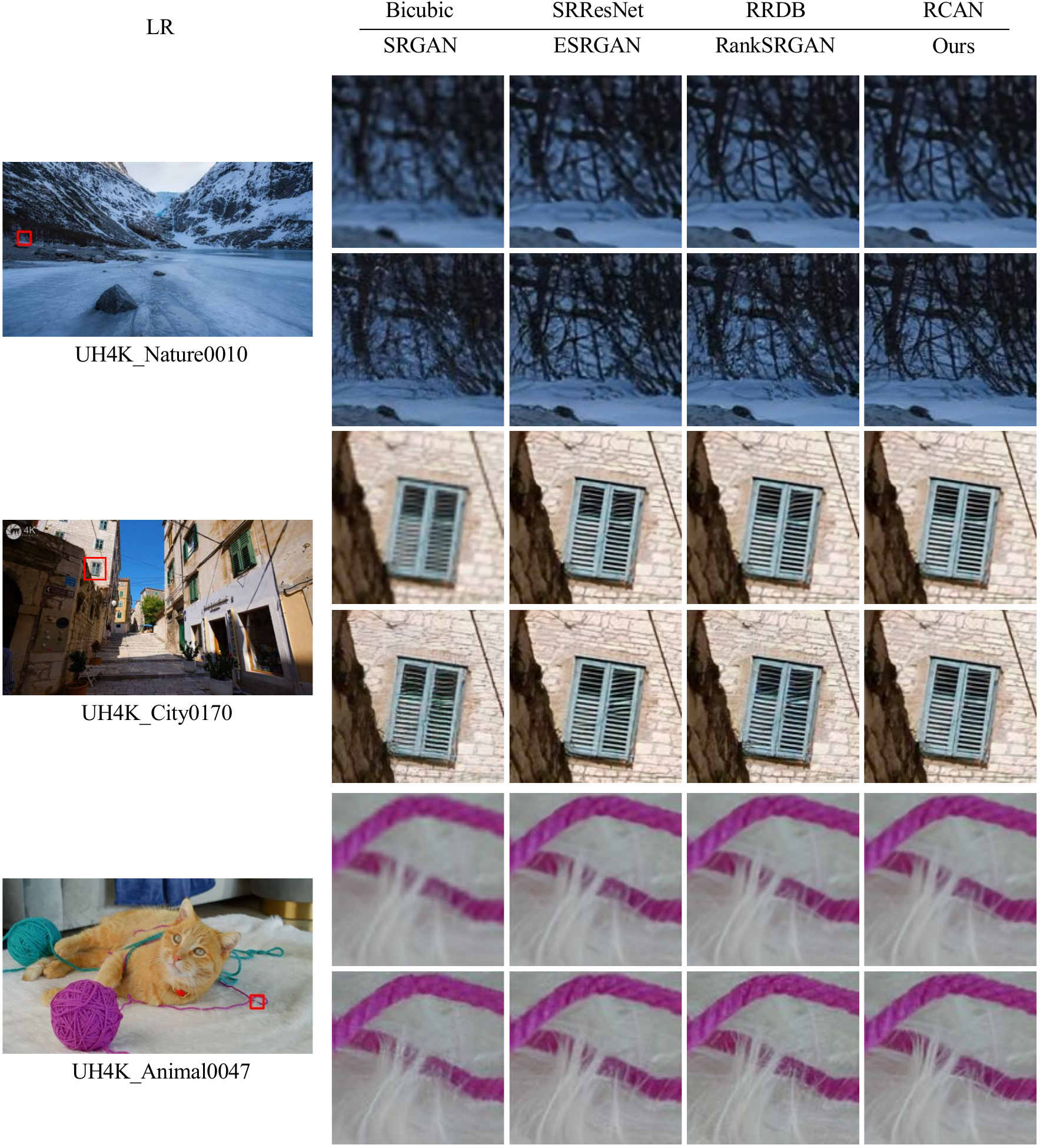}
	\end{center}
	\caption{Visual comparison of our Beby-GAN with other methods on $\times 4$ scale (part 2).}
	\label{fig:vis2}
\end{figure*}

\begin{figure*}[t]
	\begin{center}
		\includegraphics[width=1.0\linewidth]{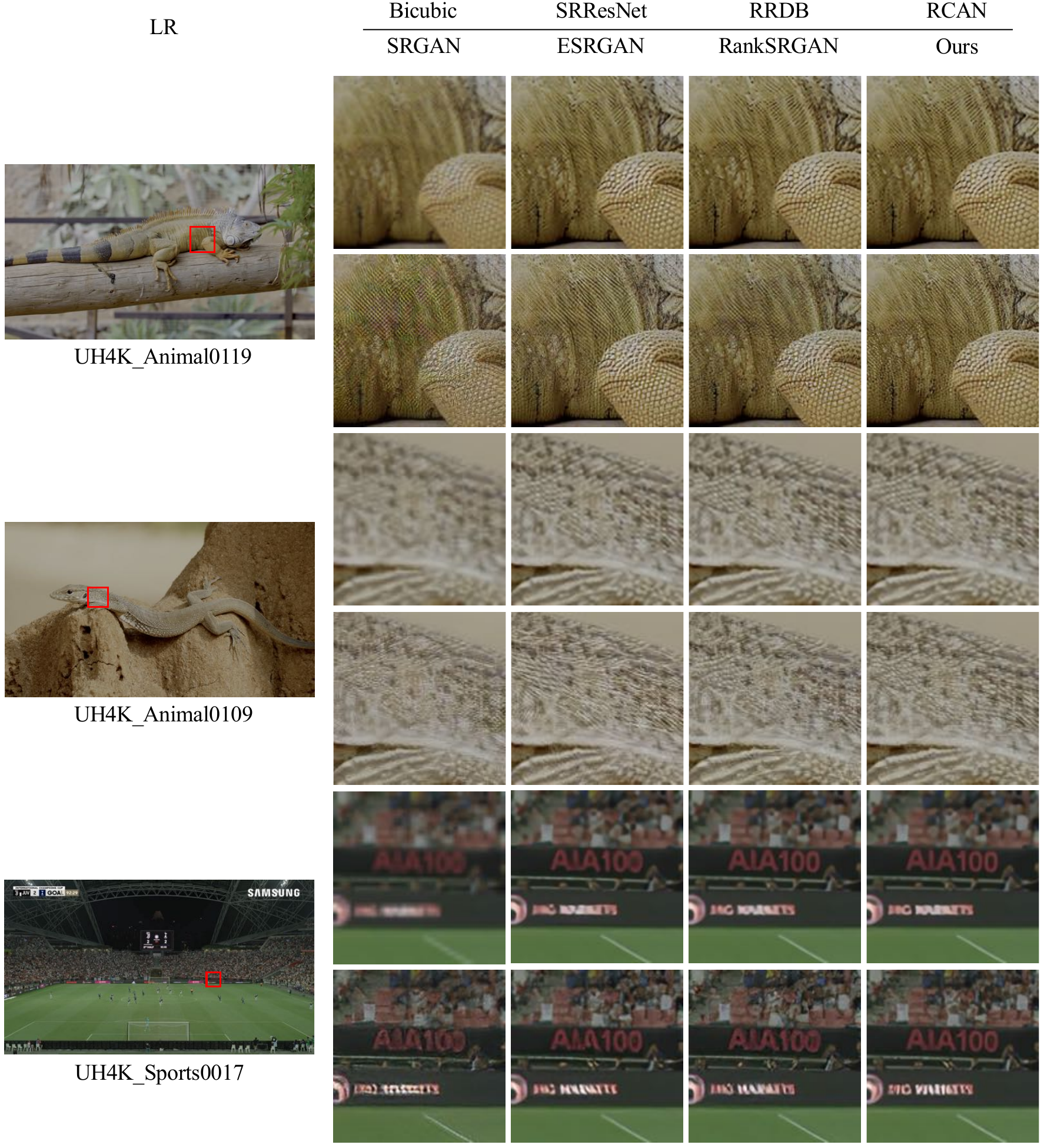}
	\end{center}
	\caption{Visual comparison of our Beby-GAN with other methods on $\times 4$ scale (part 3).}
	\label{fig:vis3}
\end{figure*}

\begin{figure*}[t]
	\begin{center}
		\includegraphics[width=1.0\linewidth]{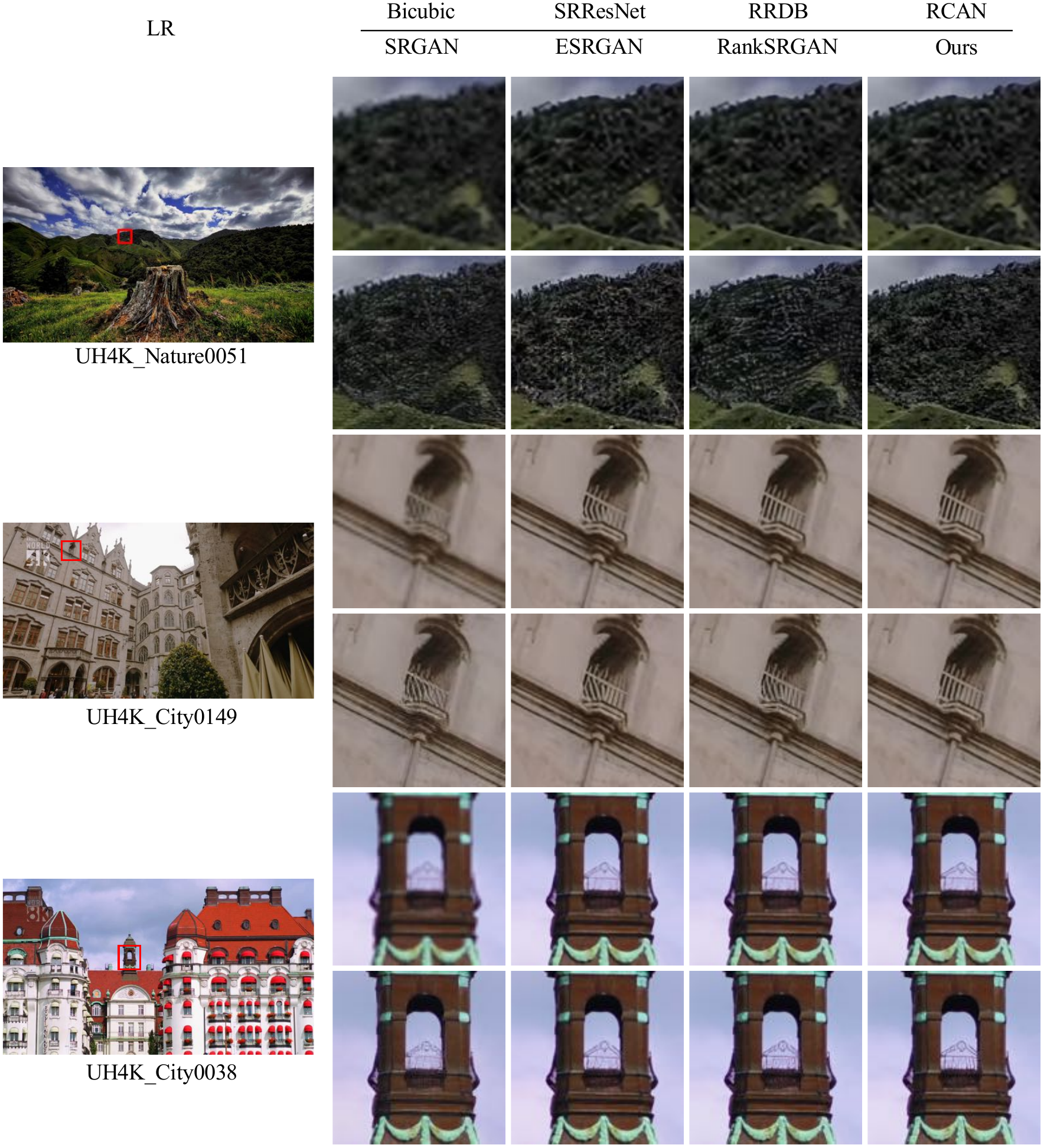}
	\end{center}
	\caption{Visual comparison of our Beby-GAN with other methods on $\times 4$ scale (part 4).}
	\label{fig:vis4}
\end{figure*}

\begin{figure*}[t]
	\begin{center}
		\includegraphics[width=1.0\linewidth]{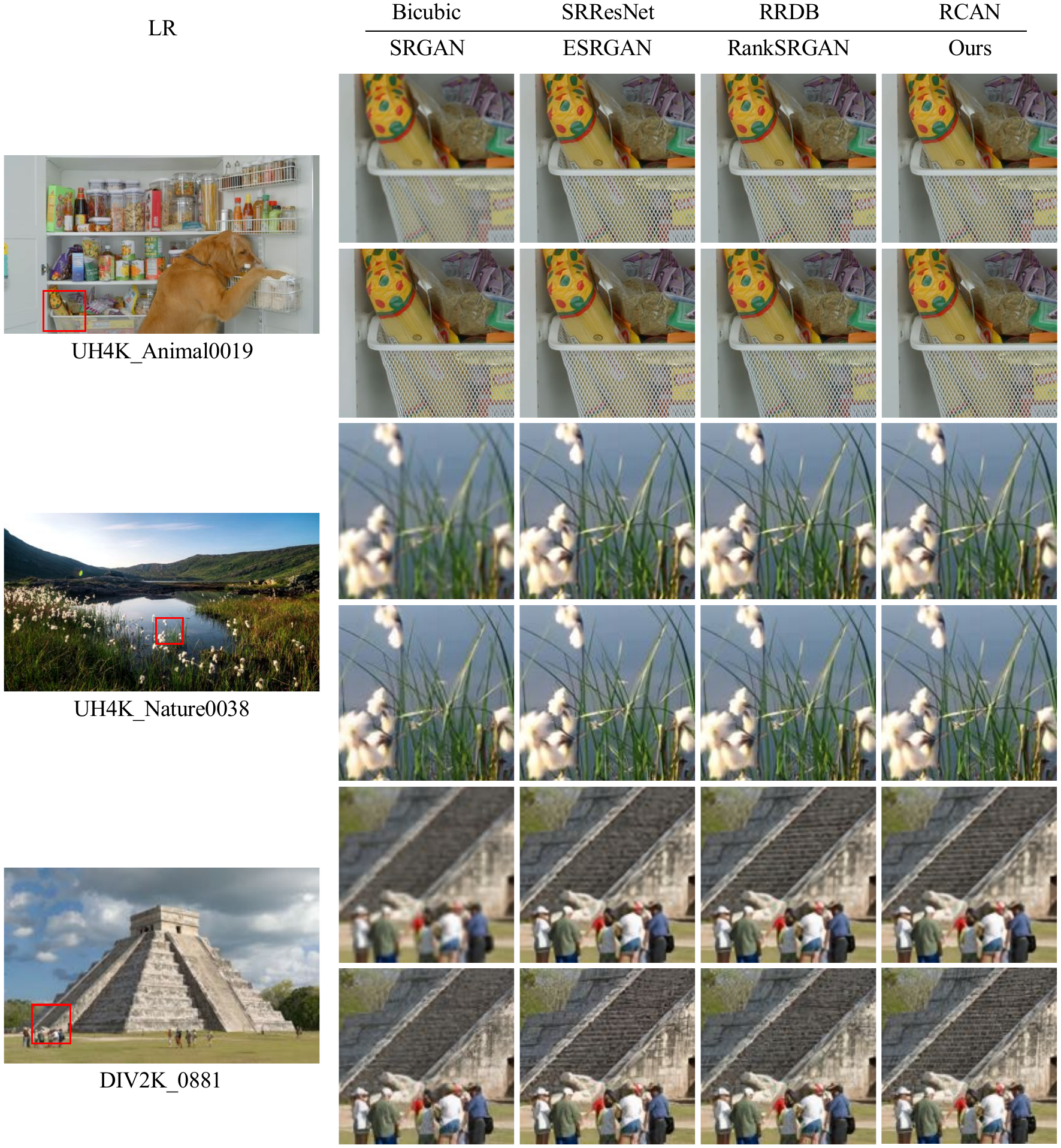}
	\end{center}
	\caption{Visual comparison of our Beby-GAN with other methods on $\times 4$ scale (part 5).}
	\label{fig:vis5}
\end{figure*}

\begin{figure*}[t]
	\begin{center}
		\includegraphics[width=1.0\linewidth]{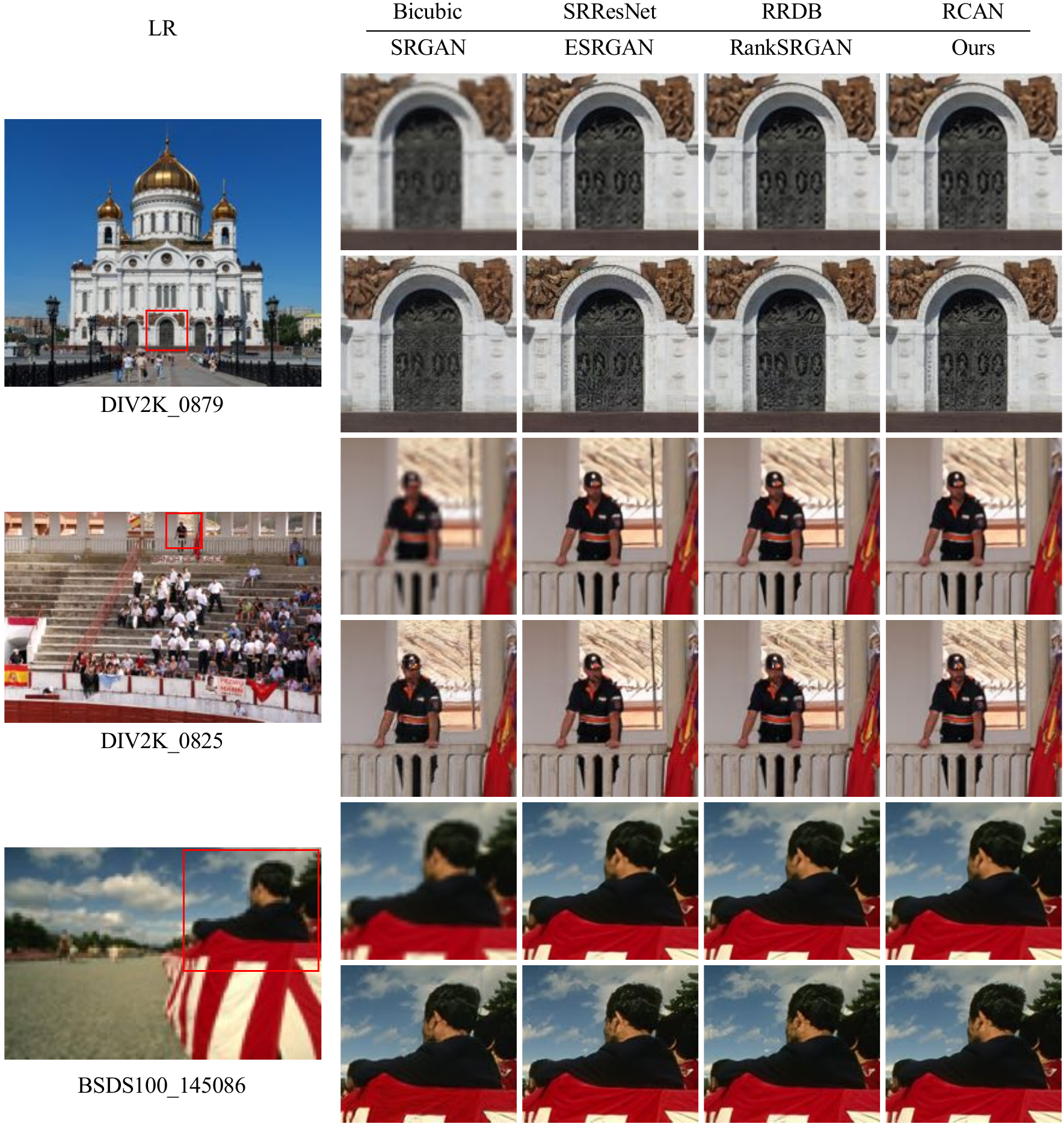}
	\end{center}
	\caption{Visual comparison of our Beby-GAN with other methods on $\times 4$ scale (part 6).}
	\label{fig:vis6}
\end{figure*}

Here we present more visual examples on our proposed ultra-high resolution 4K (UH4K), DIV2K~\cite{agustsson2017ntire} and BSDS100~\cite{martin2001database} datasets, as shown in Figure~\ref{fig:vis1}, ~\ref{fig:vis2}, ~\ref{fig:vis3}, ~\ref{fig:vis4}, ~\ref{fig:vis5}, ~\ref{fig:vis6}. Compared with SRResNet~\cite{ledig2017photo}, RRDB~\cite{wang2018esrgan}, RCAN~\cite{zhang2018image}, SRGAN~\cite{ledig2017photo}, ESRGAN~\cite{wang2018esrgan} and RankSRGAN~\cite{zhang2019ranksrgan}, our best-buddy GANs (Beby-GAN) recovers more realistic texture while maintaining the naturalness of images. Especially, it is observed that our Beby-GAN recovers more regular structures and richer details on the UH4K dataset, achieving higher perceptual quality. We also show a demo video, which compares the results of our Beby-GAN and bicubic-upsampled (BI) results.

\begin{strip}\centering
	\includegraphics[width=0.95\linewidth]{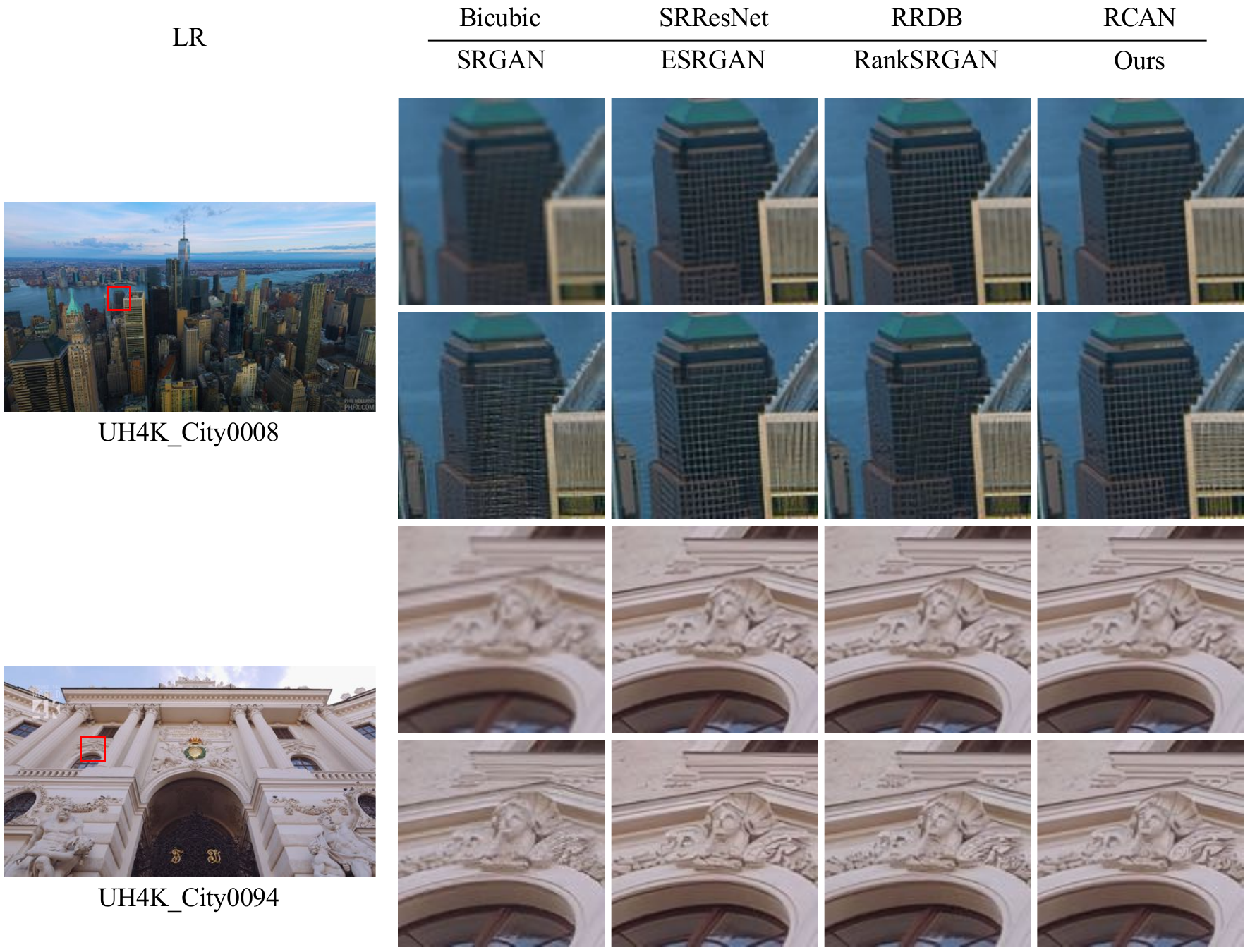}
	\captionof{figure}{Visual comparison of our Beby-GAN with other methods on $\times 4$ scale (part 1).\label{fig:vis1}} 
\end{strip}

\clearpage
\newpage


\clearpage
\newpage

\section{Masked HR Images}

\begin{strip}\centering
	\includegraphics[width=0.9\linewidth]{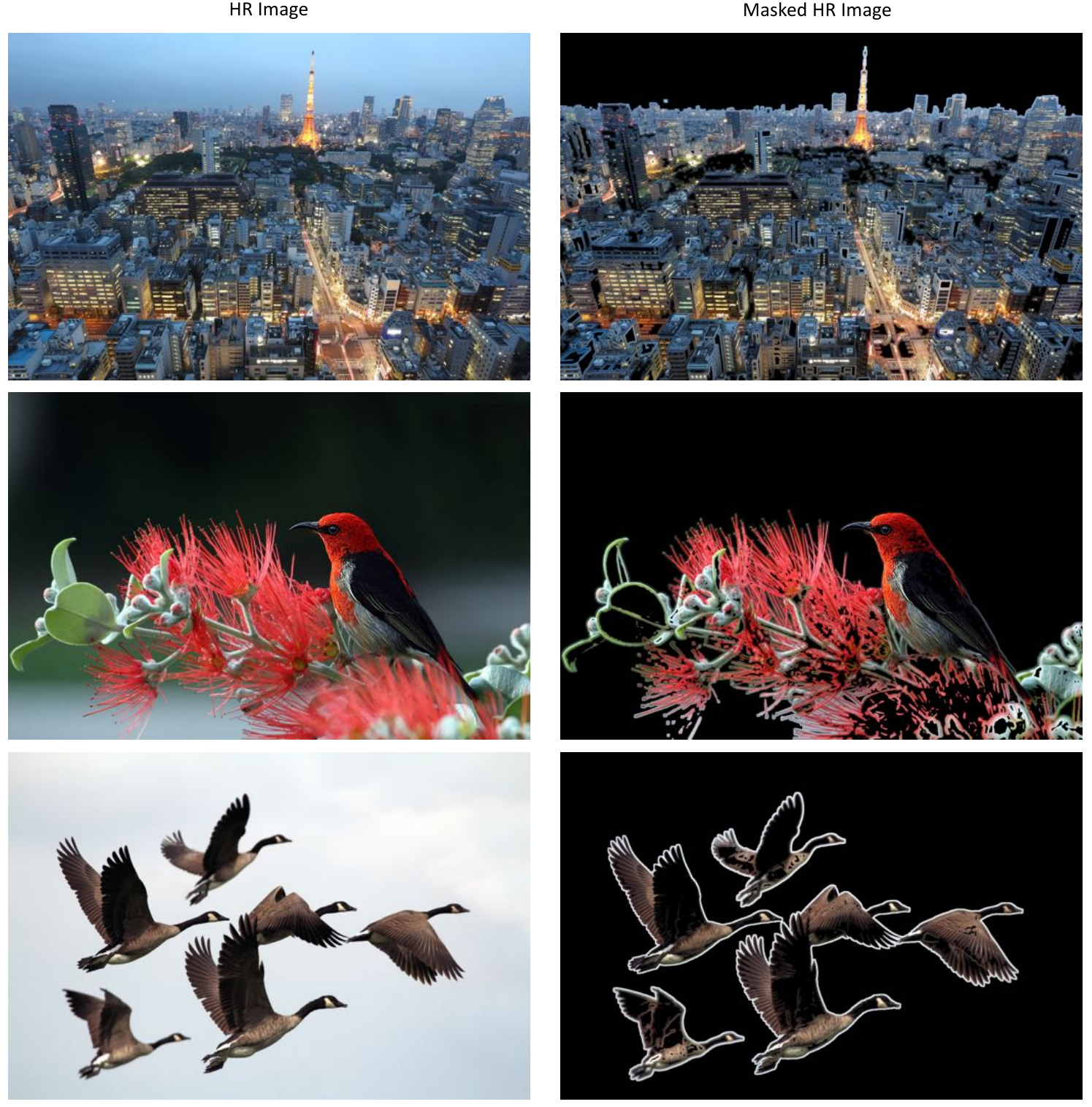}
	\captionof{figure}{Visual examples of masked HR images (part 1).\label{fig:mask1}} 
\end{strip}

\begin{figure*}[t]
	\begin{center}
		\includegraphics[width=0.9\linewidth]{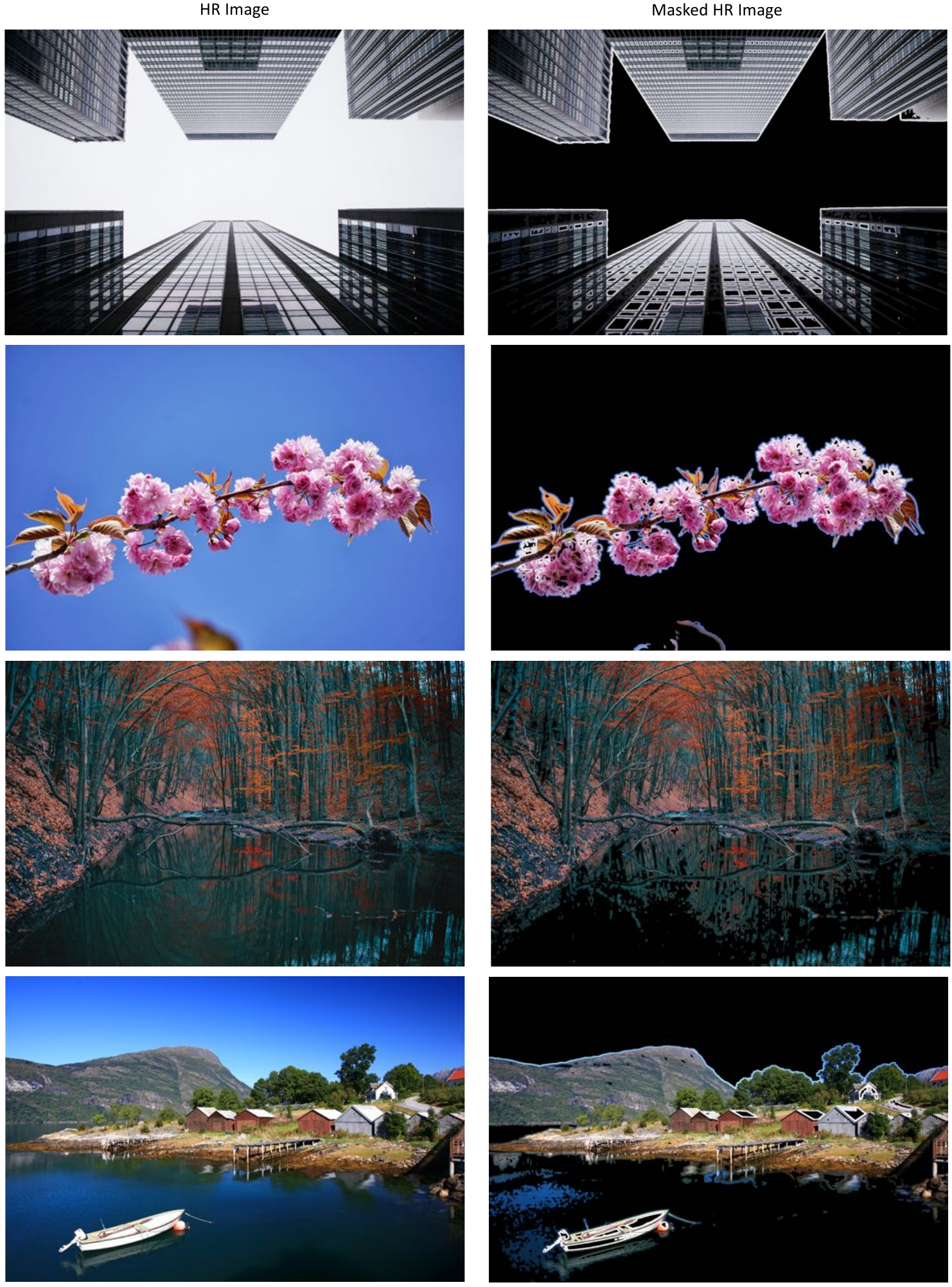}
	\end{center}
	\caption{Visual examples of masked HR images (part 2).}
	\label{fig:mask2}
\end{figure*}

\begin{figure*}[t]
	\begin{center}
		\includegraphics[width=0.88\linewidth]{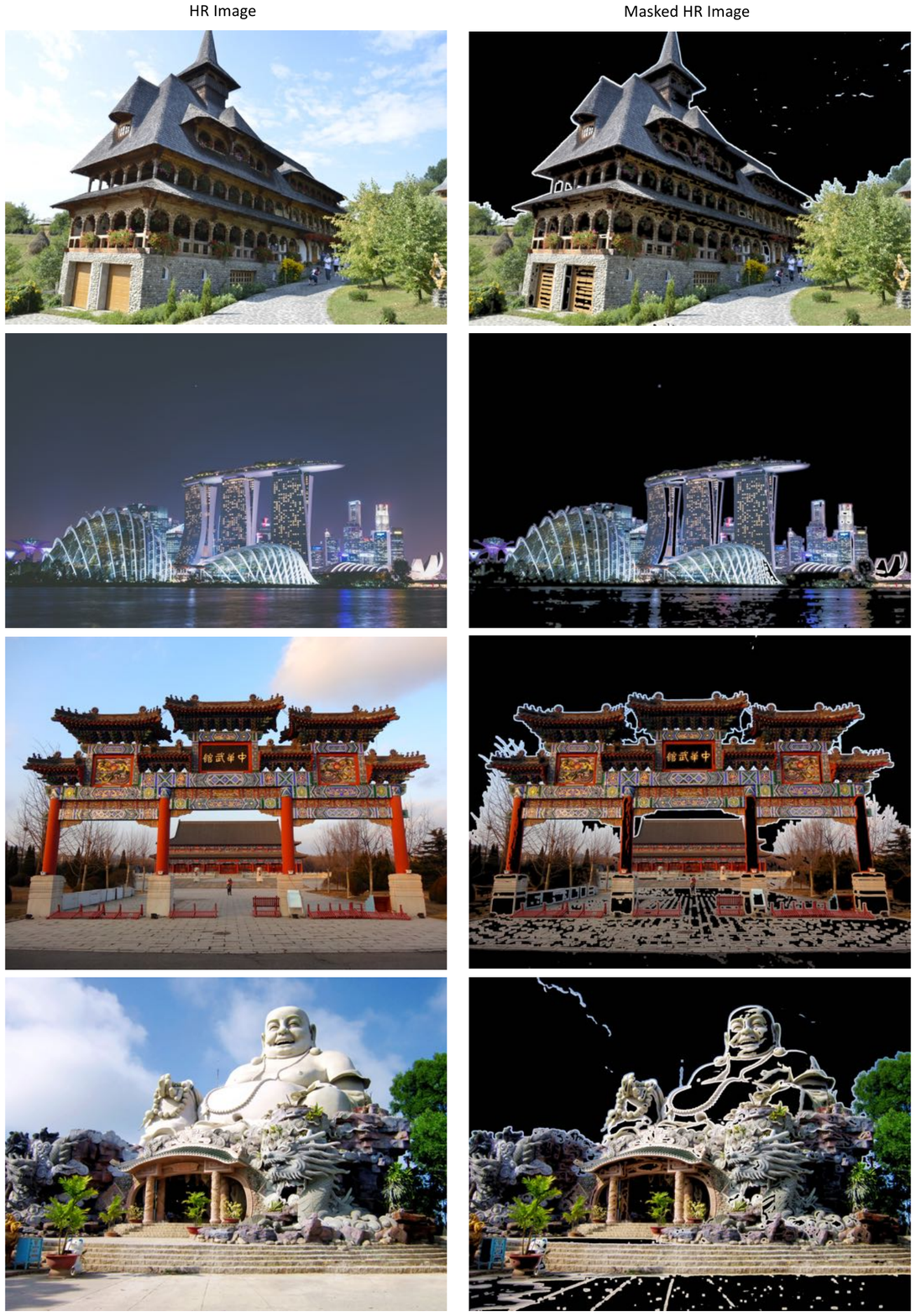}
	\end{center}
	\caption{Visual examples of masked HR images (part 3).}
	\label{fig:mask3}
\end{figure*}

In Sec.~\textcolor{red}{3.2}, we describe a region-aware adversarial learning strategy, where a mask is introduced to differentiate the rich-texture areas from smooth ones for each image. The mask is obtained from a ground-truth HR image based on Eq.~\textcolor{red}{5}. Here, we show more visual examples of HR images and the corresponding masked results in Figure~\ref{fig:mask1}, ~\ref{fig:mask2}, ~\ref{fig:mask3}. The smooth regions are masked with the black color while the rich-texture areas are preserved. It is clear that our method could separate the rich-texture and smooth regions. Though more sophisticated strategies can be used at the cost of more computation, we have demonstrated that this simple masking method already works very well, as detailed in the ablation study section.

\clearpage

\newpage




\clearpage

\section{Supplementary Results of Sect.~\textcolor{red}{4.4}}
Apart from LPIPS~\cite{zhang2018unreasonable}, PI~\cite{blau20182018}, PSNR and SSIM~\cite{wang2004image} results shown in Sect.~\textcolor{red}{4.4}, we also report the NIQE~\cite{mittal2012making} performance of our Beby-GAN-PI and Beby-GAN models in Table~\ref{tab:result}. Besides, as shown in Figure~\ref{fig:visual}, we add the ground-truth images to the visual comparison figure in Sect.~\textcolor{red}{4.4}  for a better comparison.
\vspace{0.2in}

\begin{table}[h]
	\renewcommand\arraystretch{1.1}
	\begin{center}
		\begin{tabular}{ c | c | c | c | c}
			\hline
			Datasets & Set14 & BSDS100 & DIV2K & UH4K \\
			\hline
			Beby-GAN-PI & 3.21 & 2.98 & 2.73 & 2.46 \\
			Beby-GAN & 4.09 & 3.53 & 3.21 & 3.40 \\
			\hline
		\end{tabular}
	\end{center}
	\caption{NIQE results of Beby-GAN-PI and Beby-GAN.}
	\label{tab:result}
\end{table}

%
\begin{strip}\centering
	\includegraphics[width=1.0\linewidth]{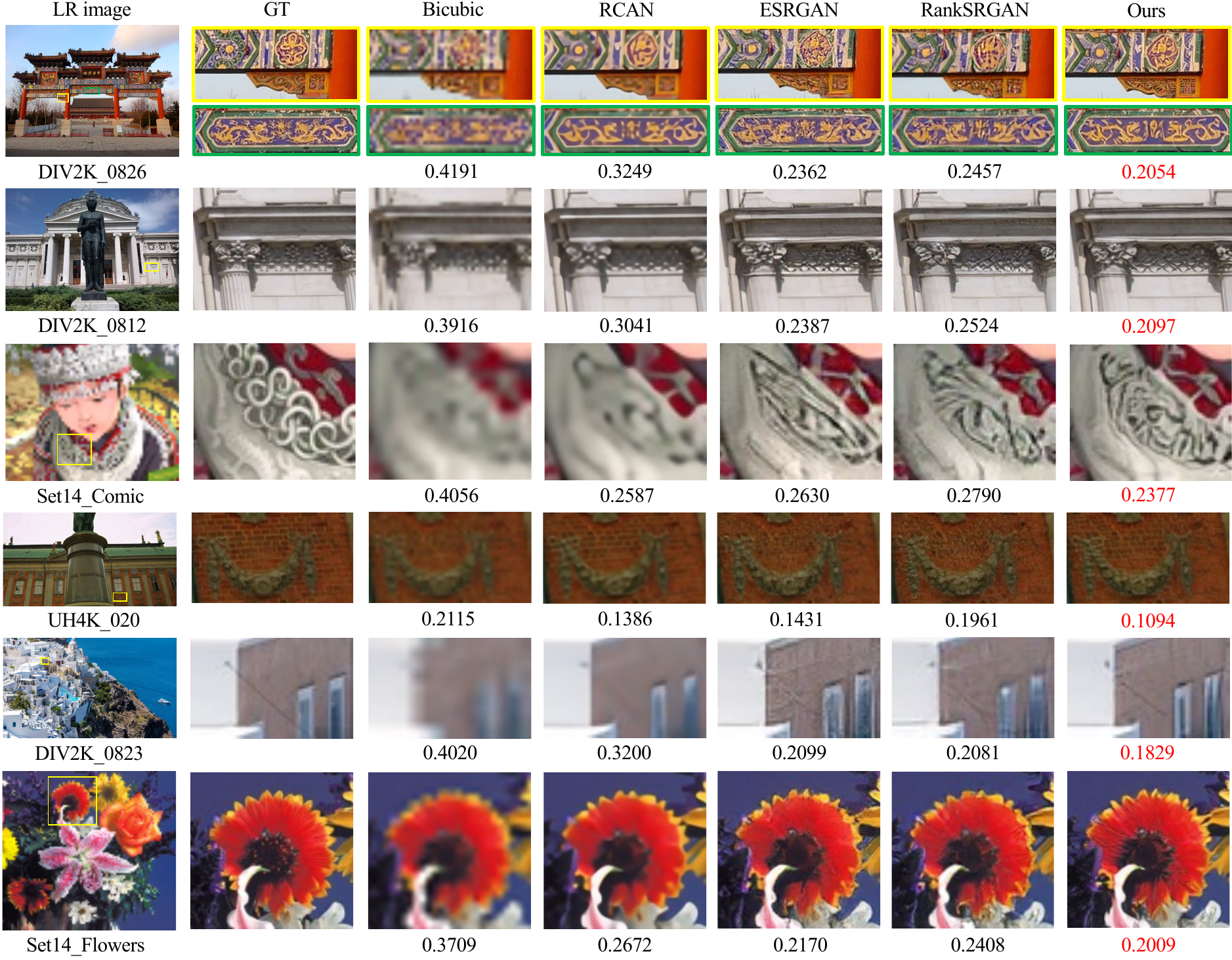}
	\captionof{figure}{Visual comparison of our Beby-GAN with other methods on $\times 4$ scale. The values beneath images represent LPIPS measures. \textcolor{red}{Red}: best quantitative results. It is clear that our method obtains the best visual performance.\label{fig:visual}} 
\end{strip}

\newpage

{\small
	\bibliographystyle{ieee_fullname}
	\bibliography{egbib}
}
\end{document}